\documentclass[twocolumn,journal]{IEEEtran}
\usepackage{amsmath,amssymb,latexsym}
\usepackage{graphicx}
\usepackage{epstopdf}
\usepackage{subfigure}
\usepackage{cite}
\usepackage{epsfig}
\usepackage{verbatim}
\usepackage{arcs}
\usepackage{yhmath}
\usepackage{setspace}
\usepackage[ruled,lined,linesnumbered]{algorithm2e}
\usepackage{pdfsync}

\newtheorem{thm}{\bf Theorem}
\newtheorem{lm}{\bf Lemma}
\newtheorem{df}{\bf Definition}

\newcounter{remark}
\newcommand{\remark}{\stepcounter{remark}\arabic{remark}}

\begin{document}

\title{Exploiting Social Trust Assisted Reciprocity (STAR) towards Utility-Optimal Socially-aware Crowdsensing}


\author{Xiaowen Gong, \IEEEmembership{Student Member, IEEE}, Xu Chen, \IEEEmembership{Member, IEEE}, Junshan Zhang, \IEEEmembership{Fellow, IEEE}, and H.~Vincent Poor, \IEEEmembership{Fellow, IEEE}
\thanks{Preliminary results of this paper has been presented in~\cite{Gong14GlobalSIP}.}
\thanks{X. Gong, X. Chen, and J. Zhang are with the School of Electrical, Computer, and Energy Engineering, Arizona State University, Tempe, AZ 85287.
Email: \{xgong9, xchen179, junshan.zhang\}@asu.edu; H.~V. Poor is with the Department of Electrical Engineering, Princeton University, Princeton, NJ 08544. Email:
poor@princeton.edu.}
\thanks{This work was supported in part by the Army Research Office under MURI Grant W911NF-11-1-0036, in part by the National Science Foundation under Grants CMMI-1435778, ECCS-1343210, CNS-1422277, ECCS-1408409, and in part by DTRA grant HDTRA1-13-1-0029.}}

\maketitle

\begin{abstract}

Mobile crowdsensing takes advantage of pervasive mobile devices to collect and process data for a variety of applications (e.g., traffic monitoring, spectrum sensing). In this study, a
socially-aware crowdsensing system is advocated, in which a cloud-based platform incentivizes mobile users to participate in sensing tasks by leveraging \emph{social trust} among users,
upon receiving sensing requests. For this system, \emph{social trust assisted reciprocity} (STAR) - a synergistic marriage of social trust and reciprocity, is exploited to design an
incentive mechanism that stimulates users' participation.

Given the social trust structure among users, the efficacy of STAR for satisfying users' sensing requests is thoroughly investigated. Specifically, it is first shown that all requests can
be satisfied if and only if sufficient \emph{social credit} can be ``transferred'' from users who request more sensing service than they can provide to users who can provide more than they
request. Then utility maximization for sensing services under STAR is investigated, and it is shown that it boils down to maximizing the utility of a \emph{circulation flow} in the combined social
graph and request graph. Accordingly, an algorithm that iteratively cancels a cycle of positive weight in the \emph{residual graph} is developed, which computes the optimal solution
efficiently, for both cases of divisible and indivisible sensing service. Extensive simulation results corroborate that STAR can significantly outperform the mechanisms using social trust
only or reciprocity only.

%

\end{abstract}


\section{Introduction}

Mobile crowdsensing has recently emerged as a promising paradigm for a variety of applications, thanks to the pervasive penetration of mobile devices to people's daily lives. Indeed, with the
development of 4G networks and powerful processors, smartphone sales crossed 1 billion units in 2013~\cite{gartner}. As smartphones are equipped with advanced sensors such as accelerometer,
compass, gyroscope, and camera, they can collectively carry out many sensing tasks, e.g., monitoring the environment. In a nutshell, by leveraging a crowd of mobile users, one can collect and
process sensed data far beyond the scope of what was possible before.

Although the benefit of crowdsensing is pronounced, performing a sensing task typically incurs \emph{overhead} for the participating user, in terms of the user's resource consumption devoted
to sensing, such as battery and computing power. Further, the participating user also incurs the risk of potential privacy loss by sharing its sensed data with others. In general, a user may
not participate in sensing without receiving adequate incentive. Therefore, effective \emph{incentive design} is essential for realizing the benefit of crowdsensing.


\begin{figure}[t]
\centering
\includegraphics[width=0.45\textwidth]{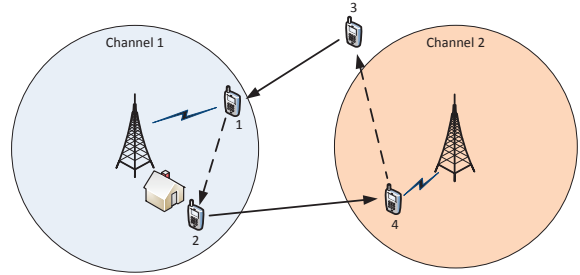}
\caption{A motivating example of social trust assisted reciprocity for spectrum crowdsensing. Users 1 and 4 have social trust in user 3 and 2, respectively (denoted by solid edges); user 2 and 3 request user 1 and 4 to sense channel 1 and 2, respectively (denoted by dashed edges). User 1 is willing to help user 2 in exchange for that user 4 helps user 1's social friend, user 3, while user 4 is willing to help user 3 in exchange for that user 1 helps user 4's social friend, user 2.}
\label{fg:cooperative_sensing}
\vspace{-0.3cm}
\end{figure}


There have been some recent studies on incentive design for crowdsensing (see, e.g., ~\cite{Yang12,Luo12,Koutsopoulos13}). Most of these works use monetary reward to stimulate users'
participation, which rely on a global (virtual) currency system. However, enforcing the circulation of a global currency typically incurs a high implementation overhead, especially for
large-scale networks, due to the need to, e.g., resolve disputes and punish counterfeiters. Therefore, it is appealing to design a crowdsensing system that can motivate a large number of
users to participate without using a global currency, which is a goal of this study.

Online social networks have been explosively growing over the past few years. In 2013, the number of online social network users worldwide reached 1.73 billion, nearly one quarter of the
world's population~\cite{emarketer}. As a result, social relationships increasingly influence people's behaviors in their interactions. In particular, as an important aspect of social
relationships, \emph{social trust} can be exploited to stimulate crowdsensing: if Alice has social trust in Bob, then Alice is willing to help Bob, since Alice can trust Bob in that Bob would
help Alice in the future to return the favor.



In this paper, we devise an incentive mechanism to stimulate users' participation in crowdsensing, by using \emph{Social Trust Assisted Reciprocity} (STAR) - a synergistic marriage of social
trust and reciprocity. Here \emph{reciprocity} means than a user helps another while it is also helped by the other. The basic idea of STAR is that Alice is willing to help Bob if someone who trusts Bob can help someone trusted by Alice. This is because the overhead of Alice for
helping Bob is compensated, as the one trusted by Alice will help Alice in the future to return the favor. We further illustrate this idea by an example of spectrum crowdsensing in
Fig.~\ref{fg:cooperative_sensing}. Without adequate incentive, user 1 and 3 are not willing to sense channel 1 and 2 for user 2 and 4, respectively. However, if user 2 and 4 are friends of
user 1 and 3, respectively, then user 1 would be willing to help user 2, since user 2's friend, user 3, will help user 1's friend, user 4, in return for user 1's help. Therefore, both user 1
and 3 have incentive to help.

By taking advantage of reciprocity (``synchronous exchange'') with the assistance of social trust (``asynchronous exchange''), STAR can efficiently encourage users' participation in
crowdsensing. In particular, STAR greatly enhances the chance that sensing requests are matched, since they can be matched through existing social trust among users. As illustrated in
Fig.~\ref{fg:cooperative_sensing}, without using either social trust or reciprocity, neither of user 1 and user 3 would be willing to help user 2 and user 4, respectively. If users are well connected in
the social network, the number of requests that can be matched with the assistance of social trust can be significant. Furthermore, compared to traditional currency-based schemes, STAR can
incur much lower implementation overhead due to the use of the already existing social trust. We will discuss the overhead of STAR and related work in Section~\ref{sc:related}.

The main thrust of this study is devoted to characterizing the fundamental performance of STAR, particularly for satisfying users' sensing requests given the social trust structure among
them. Since sensing requests are mismatched in general and social trust levels are limited, it may not be possible to satisfy all requests. Therefore, a natural question is \emph{``Can all
requests be satisfied?''} The benefit of sensing service provided under STAR can be quantified by the utility of users who receive the service. In the case that not all requests can be
satisfied, another important question arises: \emph{What is the maximum utility that can be achieved by the provided service?} These two questions are similar in spirit to admission control
and network utility maximization, respectively.


We summarize the main contributions of this paper as follows.

\begin{itemize}
\item We propose a socially-aware crowdsensing system that stimulates users' participation by leveraging their social trust. The incurred overhead can be significantly lower than that of
    traditional currency-based schemes, since it uses \emph{social credit} as a ``local'' currency enabled by social trust, rather than a global currency.

\item For the proposed system, we design STAR, an incentive mechanism which stimulates users' participation by using \emph{social trust assisted reciprocity}. We investigate thoroughly the
    efficacy of STAR for satisfying users' sensing requests, given the social trust structure among users. Specifically, we first show that all requests can be satisfied if and only if
    users who request more sensing service than they can provide can transfer sufficient social credit to users who can provide more than they request. Then we investigate utility
    maximization for sensing service, and show that this problem is equivalent to maximizing the utility of a \emph{circulation flow} in the combined social graph and request graph. Based
    on this observation, we develop an algorithm that iteratively cancels the cycles of positive weights in the \emph{residual graph}, and hence computes the optimal solution efficiently,
    for both cases of divisible and indivisible service.

\item We evaluate the performance of STAR through extensive simulations for a random setting based on the Erd\H{o}s-R\'{e}nyi graph model, and for a practical setting based on real social data with application to spectrum crowdsensing. For both settings, simulation results demonstrate that STAR can achieve significantly better system efficiency and individual user performance than only using social trust or reciprocity.

\end{itemize}


The rest of this paper is organized as follows. In Section~\ref{sc:model}, we propose a socially-aware crowdsensing system. In Section~\ref{sc:description}, we design an incentive mechanism
based on social trust assisted reciprocity for the proposed system. Based on STAR, Section~\ref{sc:efficiency} investigates conditions for satisfying all sensing requests and the utility
maximization for sensing service. Section~\ref{sc:simulation} provides simulation results to illustrate the efficacy of STAR. Related work is reviewed in Section~\ref{sc:related} and the
paper is concluded in Section~\ref{sc:conclusion}.



\section{Socially-aware Crowdsensing System}\label{sc:model}

In this section, we describe a crowdsensing system that stimulates users' participation by leveraging their social trust.

\begin{figure}[t]
\centering
\includegraphics[width=0.50 \textwidth]{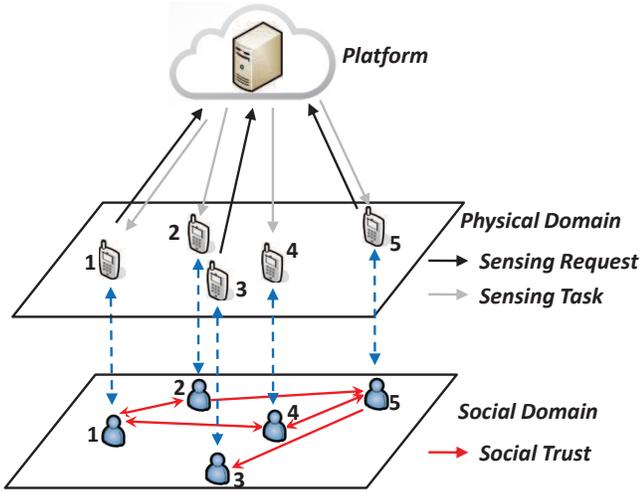}
\caption{Illustration of socially-aware crowdsensing system.}
\label{fg:cloud_platform}
\end{figure}

\subsection{Motivation}


Social relationships play an increasingly important role in people's interactions with each other. One important aspect of the social relationship between two users is their \emph{social
trust}: one user has belief in and relies on the other user's behavior in the future. To stimulate users' participation in crowdsensing, social trust can be exploited in the form of
\emph{social credit}. Specifically, social credit is transferred between two users with social trust if one user owes a favor to the other and commits to return the favor later. Therefore, a
user is willing to participate if it receives social credit from another user that it has social trust in. This ``asynchronous exchange'' of favors via social credit is in the same spirit as
a global currency. However, since this pairwise credit commitment is enabled by existing social trust between two users, social credit would incur a much lower overhead than a global
currency.

Most existing crowdsensing systems assume that a platform announces sensing tasks and motivates users to participate in these tasks by providing monetary incentives
\cite{Yang12,Koutsopoulos13}. In contrast, we are interested in a system where sensing requests are generated by users. Indeed, a few crowdsensing systems that have been deployed are based on
this model. For example, the Waze~\cite{waze} system employs traffic monitoring data collected from a crowd of drivers to answer an individual driver's request (e.g., navigation to a specific destination).


\subsection{System Description}\label{sc:model:description}

We consider a crowdsensing system as illustrated in Fig.~\ref{fg:cloud_platform}. The system consists of a \emph{platform} that operates in the cloud, and a set of mobile \emph{users} $V =
\{1,\cdots,N\}$ connected to the platform via the cloud. Initially, each user registers at the platform and publishes its social information (e.g., Facebook account) such that users can
identify their social relationships with each other. Then, each user declares to the platform a social credit \emph{limit} for each other user that it has social trust in, based on the
strength of their social relationship. The social credit limit quantifies the social trust level by specifying how much social credit one user is willing to accept from another. For example,
a user typically has high social trust in a close friend, while it may have low social trust in an acquaintance. The system proceeds in rounds and the workflow in each round consists of four
major components as depicted in Fig. ~\ref{fg:workflow}. We describe each component in detail below.

\begin{itemize}

\item \textbf{Sensing request formation.} A user can submit to the platform a \emph{sensing request} that describes the sensing service it needs. For example, a user may request to know if
    a licensed channel is available. Upon receiving a request, the platform can find a particular set of users who can serve the request, based on users' sensing capabilities, such as their
    physical locations and the functions of sensors on their devices. In this way, the platform determines the request relationships among users, i.e., which user requests service from
    which other users. For example, a user with a good sensing channel condition for a licensed channel can serve another user's request to sense that channel.


\begin{figure}[t]
\centering
\includegraphics[width=0.48 \textwidth]{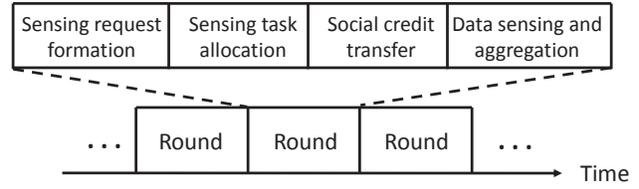}
\caption{Workflow of socially-aware crowdsensing system.}
\label{fg:workflow}
\end{figure}

\item \textbf{Sensing task allocation.} Based on the sensing requests, the platform allocates sensing tasks to users. A sensing task specifies how much sensing service is needed from that
    user. For example, a sensing task may require a user to sense a licensed channel for a period of time. A key challenge for the platform is to ensure that users have incentive to carry
    out their allocated tasks.

As expected, a user who requests sensing service can also receive requests from others for service. Therefore, it is plausible to take advantage of \emph{direct (bilateral)} or
\emph{indirect (multi-lateral) reciprocity} (as illustrated in Fig.~\ref{fg:cooperation_cycle}(a),(b)): Alice is willing to help Bob if Bob simultaneously helps Alice. While this
``synchronous exchange'' of favors is appealing as it obviates the need for currency, a major drawback is that users' requests have to be simultaneously matched, which does not hold in
general. As illustrated by the example in Fig.~\ref{fg:cooperative_sensing}, user 1 has a good sensing channel for channel 1 while user 2 does not. Therefore, user 2 needs user 1's help
while user 1 does not need user 2's help.

%

\item \textbf{Social credit transfer.} The platform can stimulate users' participation by using social credit. The platform maintains the social credit limit for each pair of users, and
    updates it for the next round to reflect the amount of social credit transferred between them in the current round. Besides the update performed by the platform, each user can also
    change its credit limit for another by reporting the new value to the platform.

\item \textbf{Data sensing and aggregation.} Based on the transferred social credit, users have incentive to carry out their allocated sensing tasks. After collecting and processing the
    sensing data from users, the platform distributes the aggregated data to the intended users. Therefore, most of the communication and computation burdens shifts from the users to the
    platform.


\end{itemize}


Unlike most existing work, our proposed crowdsensing system exploits social trust to stimulate users' participation, which obviates the need of a global currency. For this system, one key
challenge is to make the best use of social credit such that users have incentive to carry out sensing tasks, and more importantly, the system achieves good performance, which is the focus in
the rest of this paper.

\section{STAR: Social Trust Assisted Reciprocity Based Incentive Mechanism}\label{sc:description}

In this section, we design an incentive mechanism based on social trust assisted reciprocity.

\subsection{System Model}


\begin{figure}[t]
\centering
\includegraphics[width=0.35 \textwidth]{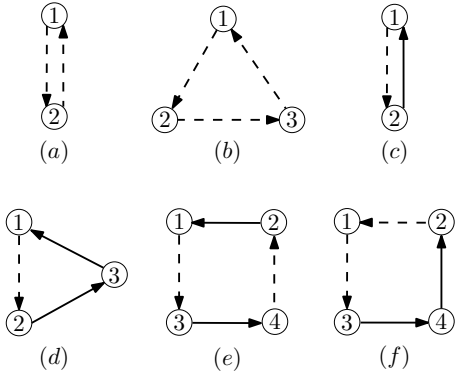}
\caption{Examples of social trust assisted reciprocity cycles. (a)-(d) are special cases: (a) direct reciprocity cycle; (b) indirect reciprocity cycle; (c) direct social trust based cycle; (d) indirect social trust based cycle. Solid edges are social edges and dashed edges are request edges.}
\label{fg:cooperation_cycle}
\end{figure}

We model users' sensing requests by a \emph{request graph} $G^R \triangleq (V, E^R)$, in which user $i$ and user $j$ are connected by a directed \emph{request edge} $e^R_{ij} \in E^R$ if user
$j$ requests sensing service\footnote{For brevity, we use ``sensing service'' and ``service'' interchangeably throughout the paper.} from user $i$. The capacity $R_{ij}>0$ of each request
edge $e^R_{ij}$ represents the \emph{amount} of service requested by user $j$ from user $i$\footnote{Recall that users' request relationships are determined by the platform in the sensing
request formation phase as described in Section \ref{sc:model:description}.}. The flow $f^R_{ij}>0$ on the request edge $e^R_{ij}$ represents the amount of service provided by user $i$ to
user $j$. Each unit of service captures a unit of sensing cost (e.g., energy consumption, privacy loss) incurred for the user who provides the service. Depending on the specific application, sensing service can be \emph{divisible} (e.g., quantified by sensing time) such that $R_{ij}$ and $f^R_{ij}$ for each $e^R_{ij} \in E^R$ have
continuous values, or \emph{indivisible} (e.g., quantified by the number of sensing data samples) such that they have to be integers. In some situations, a user cannot provide all the service
requested from it (e.g., due to its resource constraints). To take this into account, let $C_{i}$ be the maximum amount of service that user $i$ can provide and $E^C_{i}$ be the set of
outgoing request edges of user $i$. Then the following constraint applies:
\begin{align}
&\sum_{j:e_{ij}\in E^R_{i^+}} f^R_{ij} \le C_{i}. \label{capacity3}
\end{align}
We will discuss how to capture constraint~\eqref{capacity3} in our incentive mechanism in Section~\ref{sc:efficiency}.

A user obtains utility from its requested sensing service, which depends on the amount of service provided by each user who is requested for that service. We assume that user $j$ obtains a
utility of $U_{ij}$ for each \emph{unit} of service provided by user $i$. In general, user $j$ can request different types of service from user $i$, which have utilities of
different values. For example, a user may request sensing multiple channels from another user, whose sensing capability varies across different channels. In this case, there are multiple
\emph{parallel} request edges (in the same direction) from user $i$ to user $j$, each with a specific utility of service\footnote{For brevity, we say ``utility of service'' instead of
``utility per unit service''.}. In this paper, we assume that there exists \emph{at most one} request edge from one user to another. However, all the results obtained under this assumption can be directly extended to the case of parallel request edges. We further assume that a user's utility is equal to the total utility of the service provided to that user. More complex forms of utility will be studied in our future work.

We model the social trust structure among users by a \emph{social graph} $G^S \triangleq (V, E^S)$, in which user $i$ and user $j$ are connected by a directed \emph{social edge} $e^S_{ij} \in E^S$ if user $j$ has social trust in user $i$. The capacity $S_{ij}> 0$ of each social edge $e^S_{ij}$ represents the social credit limit, which specifies the maximum amount of social credit that can be transferred from user $i$ to user $j$. The flow $f^S_{ij}$ on the social edge $e^S_{ij}$ represents the amount of social credit transferred between user $i$ and user $j$. The social credit unit is the same as the sensing service unit, and is the same for all users. Note that $f^S_{ij} = -f^S_{ji}$ holds for each pair of social edges between two users, where $f^S_{ij}> 0$ (or $f^S_{ji}>0$) indicates that a credit of $f^S_{ij}$ (or $f^S_{ji}$, respectively) is transferred from user $i$ to user $j$ (or from user $j$ to user $i$, respectively).


\subsection{An Example of Spectrum Crowdsensing}\label{sc:model:example}

As an illustrative example, we next discuss how the system model described above can be applied to spectrum crowdsensing.

Spectrum sensing is an important and challenging task in cognitive radio networks \cite{Sahai09}. To access a licensed channel in a cognitive radio network, a user needs to sense the channel to ensure that the channel is not used by licensed transmitters. When a user's sensing channel condition is impaired by severe fading (e.g., path loss, shadowing), the user needs other users' help to sense the channel. Consider a cognitive radio network where each user intends to sense one or multiple licensed channels for access. A user's sensing capability for a channel depends on its sensing channel condition, which can vary across different users and different channels. If user $i$ has a good sensing channel condition for a channel, user $j$ may request user $i$ to sense that channel. The overhead incurred by sensing that channel can be user $i$'s resource consumption (e.g., device battery) for the sensing task. Therefore, the amount of sensing service $f^R_{ij}$ provided by user $i$ to user $j$ can be quantified by user $i$'s sensing time. The utility $U_{ij,k}$ of user $j$ derived from the service provided by user $i$ on channel $k$, can depend on user $i$'s sensing capability on channel $k$ as well as user $j$'s utilization efficiency (e.g., transmission rate) of channel~$k$.



\subsection{Design Description}\label{sc:description:design}

\begin{figure}[t]
\centering
\includegraphics[width=0.45 \textwidth]{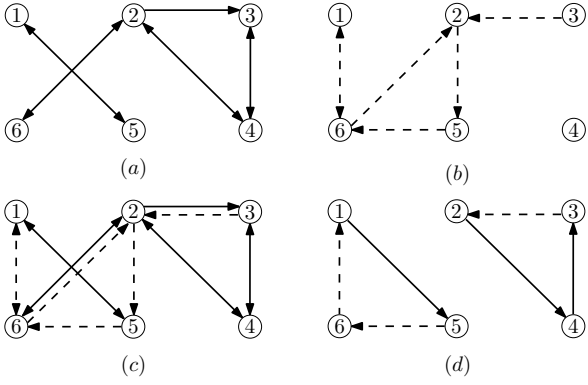}
\caption{An example of (a) social graph; (b) request graph; (c) the combined social and request graph; (d) two STAR cycles in the social-request graph.}
\label{fg:social_cooperation_graph}
\end{figure}

The basic incentive structure of the STAR mechanism is a \emph{social trust assisted reciprocity cycle} (STAR) in which a set of users have incentive to provide service. It is defined in the
combined social and request (social-request) graph $G \triangleq (V, E^S\cup E^R)$ (as illustrated in Fig.~\ref{fg:social_cooperation_graph}).
\begin{df}\label{df:cycle1}
A social trust assisted reciprocity cycle is a directed cycle in the social-request graph~$G$.
\end{df}

In a STAR cycle, a user is willing to provide service since \emph{the overhead is compensated by receiving credit or service from another user in that cycle}. For example, user 1 in
Fig.~\ref{fg:cooperation_cycle}(e) is willing to provide service to user 3 since it receives credit from user 2; user 1 in Fig.~\ref{fg:cooperation_cycle}(f) is willing to provide service to
user 3 since it receives service from user 2. Note that a STAR cycle can involve \emph{intermediate} users that only transfer credits with their social neighbors. For example, in
Fig.~\ref{fg:cooperation_cycle}(f), user 4 is an intermediate user.

For each user in a STAR cycle, the amount of service or credit it receives should be \emph{equal} to that of service or credit it provides or spends, respectively. Let $f_c$ denote a
\emph{balanced} flow along a STAR cycle $c$, which has the same flow value on each edge in $c$. The flow on a social or request edge in the \emph{aggregate} flow $f$ of a set of balanced
flows $\{f_c$, $c\in\mathcal{C}\}$ along cycles $\mathcal{C}$ is given by
\[f^S_{ij} = \!\!\!\sum_{c\in\mathcal{C}: e^S_{ij}\in c} \!\!\! f_c - \!\!\!\sum_{c\in\mathcal{C}: e^S_{ji}\in c} \!\!\!f_c, \ \ f^R_{ij} = \!\!\! \sum_{c\in\mathcal{C}: e^R_{ij}\in c} \!\!\!f_c\]
respectively. Note that the credit transferred from user $i$ to $j$ (i.e., the flow on $e^S_{ij}\in E^S$) in the balanced flow along a STAR cycle can be partly or completely \emph{canceled}
by that from user $j$ to $i$ in another STAR cycle. Users can participate in a set of balanced flows along STAR cycles if and only if the aggregate flow satisfies the capacity constraints on
request and social edges.
\begin{df}\label{df:executable}
A set of balanced flows along STAR cycles is feasible if the aggregate flow satisfies the following capacity constraints:
\begin{align}
-S_{ji} \le f^S_{ij} \le S_{ij}, &\ f^S_{ji} = - f^S_{ij}, \ \forall e_{ij} \in E^S  \label{capacity1} \\
0 \le f^R_{ij} \le &R_{ij}, \ \forall e_{ij} \in E^R. \label{capacity2}
\end{align}
\end{df}

\begin{figure}[t]
\centering
\includegraphics[width=0.32 \textwidth]{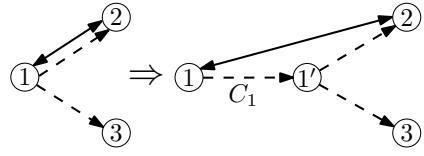}
\caption{An example of modifying a social-request graph to capture the constraint~\eqref{capacity3} with $E^R_{1^+} = \{e^R_{12},e^R_{13}\}$. The notation next to an edge is its capacity.}
\label{fg:node_capacity}
\end{figure}

Recall that the amount of service a user can provide can be constrained (i.e., constraint~\eqref{capacity3}). To capture this constraint, we can modify the social-request graph $G$ as
follows. We first construct a virtual node $i'\in V$ and change all the outgoing request edges from node $i$ to being from node $i'$, and then we add a virtual edge $e^R_{ii'}\in E^R$ and set
its capacity and utility as $C_{i}$ and 0, respectively (as illustrated in Fig.~\ref{fg:node_capacity}). Note that all other edges keep unchanged. It can be easily shown that it suffices to
focus on the modified graph: any feasible set of balanced flows along STAR cycles in the modified graph has a \emph{one-to-one correspondence} in the original graph that also satisfies
constraint~\eqref{capacity3}.


Under the STAR mechanism, all users are willing to participate in any feasible set of balanced flows along STAR cycles.

\section{Exploiting STAR to Satisfy Sensing Requests}\label{sc:efficiency}

In this section, we characterize the efficacy of the STAR mechanism. We first investigate conditions under which all sensing requests can be satisfied. Then we study the maximum total utility
that can be achieved by provided sensing service.

\subsection{Satisfying All Sensing Requests}\label{sc:efficiency:max_request}

Based on the STAR cycles, we first show that it suffices to focus on \emph{circulation} flows in the social-request graph defined as follows.
\begin{df}\label{df:circulation}
A flow $f$ in the social-request graph $G$ is a circulation if $f$ satisfies the capacity constraints~\eqref{capacity1}, \eqref{capacity2}, and the flow conservation constraints
\begin{align}
\sum_{j:e_{ij}\in E^R} \!\!\!f^R_{ij} + \!\!\!\sum_{j:e_{ij}\in E^S} \!\!\!f^S_{ij} = \!\!\!\sum_{j:e_{ji}\in E^R} \!\!\!f^R_{ji}, \ \forall i \in V. \label{conservation}
\end{align}
\end{df}

It is clear that the aggregate flow of any feasible set of balanced flows along STAR cycles is a circulation flow in $G$. The following lemma shows that the converse is also true.
\begin{lm}\label{lm:circulation1}
Any circulation flow in the social-request graph amounts to the aggregate flow of a feasible set of balanced flows along STAR cycles.
\end{lm}

\textbf{Proof}: Consider a non-empty circulation flow $f$. We can find a node $v_1$ with a positive flow on an outgoing edge from $v_1$ and trace along this edge to another node $v_2$. Due to
the flow conservation constraint, we can find an outgoing edge from $v_2$ with a positive flow and trace along it to a node $v_3$. We continue this tracing process until we visit a node $v_j$
that has been visited before, i.e., $v_i = v_j$ for some $i < j$, and hence we find a STAR cycle $v_i\rightarrow v_{i+1}\rightarrow \cdots\rightarrow v_j$. Then we subtract flow $f$ by a
balanced flow along this cycle with value equal to the minimum flow value on an edge in that cycle. Thus the remaining flow is still a circulation flow in which the number of edges with
non-zero flows is reduced. We can repeat this argument to subtract the remaining flow by a balanced flow along a cycle until it is empty. This implies that flow $f$ is the aggregate flow of
the subtracted balanced flows along the cycles, which is also feasible. $\hfill\square$

\begin{figure}[t]
\centering
\includegraphics[width=0.40\textwidth]{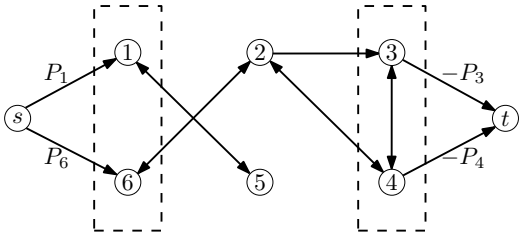}
\caption{The extended social graph constructed from the social graph in Fig.~\ref{fg:social_cooperation_graph} where $P_1 > 0$, $P_2 = 0$, $P_3 < 0$, $P_4 < 0$, $P_5 = 0$, $P_6 > 0$. The notation next to an edge is its capacity.}
\label{fg:extendend_graph}
\end{figure}

We define $P_i$ as the total amount of service requested by user $i$ deducted by the amount that user $i$ can provide:
\[P_i \triangleq \!\!\!\sum_{j:e_{ji}\in E^R}\!\!\!R_{ji} - \!\!\!\sum_{j:e_{ij}\in E^R}\!\!\!R_{ij}. \]
Then we construct an extended social graph $G^{S^+}$ from the social graph $G^S$ by adding a directed edge with capacity $P_i$ from a virtual source node $s$ to each node $i$ with $P_i > 0$,
and adding a directed edge with capacity $-P_i$ from each node $i$ with $P_i < 0$ to a virtual destination node $t$ (as illustrated in Fig.~\ref{fg:extendend_graph}). Let $P$ be defined as
\[P \triangleq \sum_{i:P_i>0} P_i = -\sum_{i:P_i<0} P_i.\]
\begin{thm}\label{thm:unbalanced_request}
All sensing requests can be satisfied under STAR if and only if $P$ is equal to the maximum flow value from $s$ to $t$ in the extended social graph $G^{S^+}$.
\end{thm}

\textbf{Proof}: By Lemma~\ref{lm:circulation1}, all requests can be satisfied if and only if there is a circulation flow $f$ in the social-request graph $G$ that saturates all request edges
(i.e., $f^R_{ij} = R_{ij}$, $\forall e^R_{ij}\in E^R$).

We first show the ``if'' part. Suppose $S$ is equal to the value of the maximum flow $f^{S^+}$ from $s$ to $t$ in $G^{S^+}$. Let $f^S$ be the flow comprised of the flows on the social edges
$E^S$ in $f^{S^+}$ (i.e., not including the edges from $s$ and to $t$ in $G^{S^+}$). Let $f^R$ be the flow in the request graph $G^R$ that saturates all request edges. Then we augment flow
$f^S$ in the social-request graph $G$ with flow $f^R$ to obtain a flow $f$ in $G$. According to the construction of $G^{S^+}$, we have $\sum_{j:e^S_{ij}\in E^S} f^S_{ij} = P_i$ for each node
$i\in V$, while we also have $\sum_{j:e^R_{ji}\in E^R} f^R_{ji} - \sum_{j:e^R_{ij}\in E^R} f^R_{ij} = P_i$. This shows that $f$ is a circulation flow.

Next we show the ``only if'' part. Suppose $f$ is a circulation flow in $G$ that saturates all request edges. Let $f^S$ be the flow comprised of the flows on the social edges $E^S$ in $f$.
Then we augment flow $f^S$ with saturated flows on the edges from $s$ and to $t$ in $G^{S^+}$ to obtain a flow $f^{S^+}$ in $G^{S^+}$. According to the construction of $G^{S^+}$, $f^{S^+}$ is
a flow in $G^{S^+}$ satisfying the capacity and flow conservation constraints, with a flow value of $P$ from $s$ to $t$. $\hfill\square$

\textbf{Remark \remark:} Theorem~\ref{thm:unbalanced_request} provides a useful insight: all requests can be satisfied if and only if \emph{users who request more service than they can provide
can transfer sufficient social credit to users who can provide more than they request, to compensate their imbalance in requests}. Intuitively speaking, the social graph serves as a
``buffer'' to partially or completely ``absorb'' the mismatch among users' requests. It is worth noting that the maximum amount of service provided under STAR is in general \emph{not} equal
to the maximum flow value from $s$ to $t$ in $G^{S^+}$.

\textbf{Remark \remark:} We note that an important difference between~\cite{Liu10} and our study is that the results in~\cite{Liu10} is based on the assumption that \emph{all users are
connected in the social network}, whereas our model here does not have this assumption. This is essentially because that reciprocity is used in STAR but not in~\cite{Liu10}. As illustrated in
Fig.~\ref{fg:cooperative_sensing}, without using reciprocity, user 1 and 4 are not willing to help user 2 and 3, respectively. In Section~\ref{sc:simulation}, simulation results demonstrate
that the STAR mechanism can significantly outperform the mechanism in~\cite{Liu10}.



\subsection{Utility Maximization for Sensing Service}\label{sc:efficiency:utility}

\begin{figure}[t]
\centering
\includegraphics[width=0.40 \textwidth]{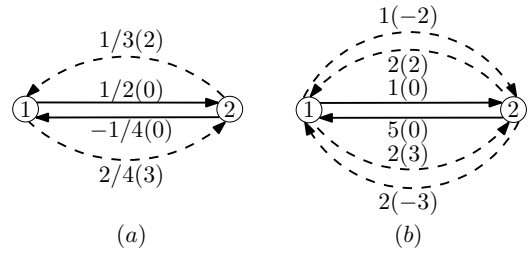}
\caption{An example of a social-request graph with a flow in (a) and its residual graph in (b). For each edge, the number before $/$ is the flow value; the number before () is the capacity; the number in () is the weight.}
\label{fg:residual_graph}
\end{figure}

Due to the mismatch of sensing service requests and social credit limits, it is possible that not all requests can be satisfied. In this case, a natural objective from the platform's view is
to maximize the total utility of provided service. The next result follows from Lemma~\ref{lm:circulation1}.
\begin{thm}\label{thm:equivalent1}
The maximum utility of sensing service provided under STAR is equal to the maximum utility of a circulation flow in the social-request graph.
\end{thm}

Note that the flow on a social edge does not generate any utility. By Theorem~\ref{thm:equivalent1}, our problem can be written as
\begin{align}
&\underset{f^S_{ij},f^R_{ij}}{\textrm{maximize}} \quad \sum_{i,j: e_{ij}\in E^R} U_{ij}f^R_{ij} \label{prb1}\\
&\textrm{subject to} \quad  \textrm{constraints} \ \eqref{capacity1}, \eqref{capacity2}, \eqref{conservation}. \nonumber
\end{align}
Note that we can \emph{maximize the total amount of service provided under STAR by solving problem \eqref{prb1} with the utility $U_{ij}$ set to 1 for each request edge}.

In the following, we will solve problem \eqref{prb1} using an algorithm inspired by the \emph{cycle-canceling} algorithm for solving the minimum cost flow problem \cite{Ahuja93}. We should
note that problem \eqref{prb1} is quite different from a typical network flow problem in that two nodes can be connected by multiple edges (request edges and social edges). Furthermore,
\emph{request edges and social edges carry different types of flows} (as illustrated in Fig.~\ref{fg:residual_graph}(a)): the flows on all request edges are \emph{non-negative} and
\emph{independent} (as in constraint~\eqref{capacity2}), while the flows on social edges can be \emph{negative} and must be \emph{inverse} between a pair of users (as in
constraint~\eqref{capacity1}).


\begin{figure}[t]
\centering
\includegraphics[width=0.45 \textwidth]{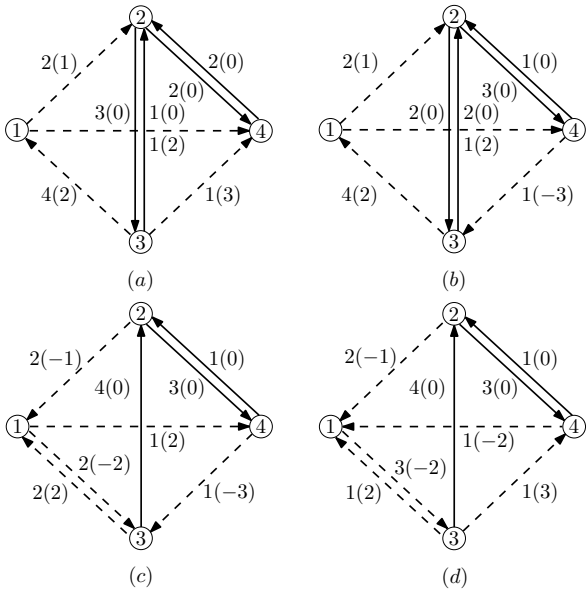}
\caption{An example of running Algorithm~\ref{al:cycle_canceling}. (a) Initial social-request graph with the empty flow; (b) Residual graph after augmenting with a flow of value 1 along cycle $2\rightarrow3\rightarrow4\rightarrow2$; (c) Residual graph after augmenting with a flow of value 2 along cycle $1\rightarrow2\rightarrow3\rightarrow1$; (d) Residual graph after augmenting with a flow of value 1 along cycle $1\rightarrow4\rightarrow3\rightarrow1$. For each edge, the number before () is the capacity; the number in () is the weight.}
\label{fg:cycle_canceling}
\vspace{-0.8cm}
\end{figure}

We start with constructing a \emph{residual graph} $G_f \triangleq (V, E^S_f \cup E^R_f)$ of the social-request graph $G$ for a given flow $f$. Specifically, for each request edge $e^R_{ij}
\in E^R$, we construct a pair of \emph{forward} edge $\overrightarrow{e}^R_{ij}\in E^R_f$ and \emph{backward} edge $\overleftarrow{e}^R_{ji}\in E^R_f$ with capacities
\[\overrightarrow{R}_{ij} = R_{ij} - f^R_{ij}, \ \overleftarrow{R}_{ij} = f^R_{ij} \]
respectively. For each \emph{pair} of social edges $e^S_{ij}, e^S_{ji} \in E^S$, we construct a pair of edges $\overrightarrow{e}^S_{ij},\overrightarrow{e}^S_{ji}\in E^S_f$ with capacities
\[\overrightarrow{S}_{ij} = S_{ij} - f^S_{ij}, \ \overrightarrow{S}_{ji} = S_{ji} - f^S_{ji} \]
respectively. We do \emph{not} construct an edge in the residual graph if its capacity is zero. Then we set the \emph{weights} of each forward edge $\overrightarrow{e}^R_{ij}\in E^R_f$ and
each backward edge $\overleftarrow{e}^R_{ji}\in E^R_f$ as
\[\overrightarrow{W}^R_{ij} = U_{ij}, \ \overleftarrow{W}^R_{ij} = -U_{ij}\]
respectively. The weights of each pair of edges $\overrightarrow{e}^S_{ij},\overrightarrow{e}^S_{ji}\in E^S_f$ are set to
\[\overrightarrow{W}^S_{ij} = \overrightarrow{W}^S_{ji} = 0.\]
We show how to construct the residual graph by an illustrative example in Fig.~\ref{fg:residual_graph}. The following lemma establishes the optimality condition for solving
problem~\eqref{prb1}.
\begin{algorithm}
\SetKwInOut{Input}{input}\SetKwInOut{Output}{output}
\Input{Social-request graph $G$}
\Output{The optimal flow for problem \eqref{prb1}} Initialize an empty flow $f$ in $G$\; \While{There exists a cycle of positive weight in the residual graph $G_f$ of flow $f$}{ Find a cycle
$c$ of positive weight in $G_f$\; Compute the residual capacity $r_c$ of cycle $c$\; Augment flow $f$ with a balanced flow of value $r_c$ along cycle $c$\; } \Return Flow $f$\; \caption{Find
the optimal flow for problem \eqref{prb1} in social-request graph $G$}\label{al:cycle_canceling}
\end{algorithm}

\begin{lm}\label{lm:optimal}
A flow $f$ is optimal for problem \eqref{prb1} if and only if there exists no cycle of positive weight in the residual graph $G_f$.
\end{lm}

\textbf{Proof}: The ``only if'' part is easy to show: If there exists a cycle of positive weight in $G_f$, then we can augment the flow $f$ with a balanced flow of value $\epsilon > 0$ along
that cycle to construct a circulation flow with larger utility.

Next we show the ``if'' part. Suppose there exists no cycle of positive weight in $G_f$ but there exists a circulation flow $f'$ in $G$ with larger utility than $f$. Similar to the residual
graph $G_f$, we construct a graph $\overline{G} \triangleq (V, \overline{E}^S\cup\overline{E}^R)$ from $G$ by constructing $\overrightarrow{e}^R_{ij},
\overleftarrow{e}^R_{ij}\in\overline{E}^R$ for each $e^R_{ij}\in E^R$ and $\overrightarrow{e}^S_{ij}, \overrightarrow{e}^S_{ji}\in\overline{E}^S$ for each pair of $e^S_{ij},e^S_{ji}\in E^S$,
and setting their weights the same as those in $G_f$. The difference between $G_f$ and $\overline{G}$ is that all the edges are constructed in $\overline{G}$ (an edge is not constructed in
$G_f$ if its capacity is 0) and have unlimited capacities. Therefore, the edges in $G_f$ is a subset of the edges in $\overline{G}$. Then we can define a flow $g$ in $\overline{G}$ by
defining the flows in $g$ on the edges of $\overline{G}$ as
\[\overrightarrow{g}^R_{ij} = \max\{0, f'^R_{ij} - f^R_{ij}\}, \ \forall \overrightarrow{e}^R_{ij}\in \overline{E}^R\]
\[\overleftarrow{g}^R_{ij} = \max\{0, f^R_{ij} - f'^R_{ij}\}, \ \forall \overleftarrow{e}^R_{ij}\in \overline{E}^R\]
\[\overrightarrow{g}^S_{ij} = f'^S_{ij} - f^S_{ij}, \ \forall \overrightarrow{e}^S_{ij}\in \overline{E}^S.\]
It follows from the definition that
\[\overrightarrow{g}^R_{ij} - \overleftarrow{g}^R_{ij} = f'^R_{ij} - f^R_{ij}, \ \forall e^R_{ij}\in E^R.\]
Then the net flow value at each node $i\in V$ in flow $g$ is
\begin{align*}
&\sum_{ j:\overrightarrow{e}^R_{ij}\in \overline{E}^R} \!\!\!\!\!\! \overrightarrow{g}^R_{ij} + \!\!\!\!\!\!\!
\sum_{ j:\overleftarrow{e}^R_{ji}\in \overline{E}^R} \!\!\!\!\!\! \overleftarrow{g}^R_{ji} + \!\!\!\!\!\!\!
\sum_{ j:\overrightarrow{e}^S_{ij}\in \overline{E}^R} \!\!\!\!\!\! \overrightarrow{g}^S_{ij} - \!\!\!\!\!\!\!
\sum_{ j:\overleftarrow{e}^R_{ij}\in \overline{E}^R} \!\!\!\!\!\! \overleftarrow{g}^R_{ij} - \!\!\!\!\!\!\!
\sum_{ j:\overrightarrow{e}^R_{ji}\in \overline{E}^R} \!\!\!\!\!\! \overrightarrow{g}^R_{ji}  \displaybreak[1]\\
&= \!\!\!\!\sum_{j: e^R_{ij}\in E^R}\!\!\!\!\!\left(f'^R_{ij} - f^R_{ij}\right) - \!\!\!\!\!\!
\sum_{j: e^R_{ji}\in E^R}\!\!\!\!\!\left(f'^R_{ji} - f^R_{ji}\right) + \!\!\!\!\!\!
\sum_{j: e^S_{ij}\in E^S}\!\!\!\!\!\left(f'^S_{ij} - f^S_{ij}\right)  \displaybreak[1]\\
&= \left(\sum_{j: e^R_{ij}\in E^R}\!\!\!\!\!f'^R_{ij} + \!\!\!\!\!\sum_{j: e^S_{ij}\in E^S}\!\!\!\!\!f'^S_{ij} - \!\!\!\!\!\sum_{j: e^R_{ji}\in E^R}\!\!\!\!\!f'^R_{ji}\right) \displaybreak[1]\\
&\hspace{1.9cm} - \left(\sum_{j: e^R_{ij}\in E^R}\!\!\!\!\!f^R_{ij} + \!\!\!\!\!\sum_{j: e^S_{ij}\in E^S}\!\!\!\!\!f^S_{ij} - \!\!\!\!\!\sum_{j: e^R_{ji}\in E^R}\!\!\!\!\!f^R_{ji}\right) = 0
\end{align*}
where the last equality follows from that $f'$ and $f$ are circulation flows in $G$. Therefore, $g$ is a circulation flow in $\overline{G}$. We observe that the flow on any edge
$e\in\overline{E}^R\setminus E^R_f$ is zero in $g$ because 1) if $e = \overrightarrow{e}^R_{ij}$, then we have $f^R_{ij} = R_{ij}$ and hence $\overrightarrow{g}^R_{ij} = 0$; 2) if $e =
\overleftarrow{e}^R_{ij}$, then we have $f^R_{ij} = 0$ and hence $\overleftarrow{g}^R_{ij} = 0$. We further observe that $\overrightarrow{g}^S_{ij} \le 0$ for any edge
$\overrightarrow{e}^S_{ij}\in\overline{E}^S\setminus E^S_f$ since we have $f^S_{ij} = S_{ij}$. Since $\overrightarrow{W}^S_{ij} = 0, \forall \overrightarrow{e}^S_{ij}\in \overline{E}^S$, the
weight of flow $g$ in $\overline{G}$ is
\begin{align*}
\sum_{i,j:\overrightarrow{e}^R_{ij}\in \overline{E}^R}\!\!\!\!\!\left(\overrightarrow{W}^R_{ij}\overrightarrow{g}^R_{ij} \right.&+\left. \overleftarrow{W}^R_{ij}\overleftarrow{g}^R_{ij}\right) = \!\!\!\!\!
\sum_{i,j:e^R_{ij}\in E^R} \!\!\!\!\! U_{ij}\left(f'^R_{ij} - f^R_{ij}\right) \\
&\hspace{0.5cm}= \!\!\!\!\!\sum_{i,j:e^R_{ij}\in E^R} \!\!\!\!\!U_{ij}f'^R_{ij} - \!\!\!\!\!\sum_{i,j:e^R_{ij}\in E^R} \!\!\!\!\!U_{ij}f^R_{ij} > 0
\end{align*}
where the last inequality follows from the assumption that $f'$ has larger utility than $f$ in $G$. Since $g$ only has positive flows on the edges in $G_f$, using a similar argument as in the
proof of Lemma~\ref{lm:circulation1}, $g$ is the aggregate flow of balanced flows along cycles each comprised of edges in $G_f$. Then the total weight of these flows along the cycles in $G_f$
is equal to the weight of flow $g$ in $\overline{G}$, which is greater than 0. This implies that there must exist a cycle of positive weight in $G_f$, which is a contradiction to the previous
assumption. This completes the proof. $\hfill\square$

Using Lemma~\ref{lm:optimal}, we can develop an algorithm as described in Algorithm~\ref{al:cycle_canceling} to solve problem \eqref{prb1}. The algorithm starts with the empty flow in the
network. It iteratively finds a cycle of \emph{positive weight} in the residual graph and cancels each cycle by augmenting the current flow in the graph with a balanced flow along that cycle,
until no cycle of positive weight exists. In each iteration, the value of the flow to augment with is set to be the \emph{residual capacity} of the cycle, which is the minimum capacity of all
edges in that cycle. We show how Algorithm~\ref{al:cycle_canceling} works by an illustrative example in Fig.~\ref{fg:cycle_canceling}.

As for the step 2 and 3 in Algorithm~\ref{al:cycle_canceling}, we can use an algorithm similar to the Bellman-Ford algorithm\cite{Huang06} to find a cycle of positive weight in the residual graph,
if there exists one. In particular, the algorithm iteratively updates the maximum weight $M(t)$ from a source node $s\in V$ to each other node $t\in V\setminus\{s\}$. In each iteration, the
algorithm checks each edge $e^S_{ij}\in E^S_f$ or $e^R_{ij}\in E^R_f$ once, and increases the maximum weight $M(j)$ to $M(i) + \overrightarrow{W}^S_{ij}$ if $M(i) + \overrightarrow{W}^S_{ij}
> M(j)$. The algorithm runs for $|V| - 1$ iterations. When it terminates, if $M(t)$ for some $t\in V\setminus\{s\}$ can be further reduced by checking some edge, then there exists a cycle of
positive weight in the graph. The algorithm has running time $O(|V|(|E^S| + |E^R|)$.


For ease of exposition, we will focus on problem \eqref{prb1} with \emph{rational} parameters: the utilities and capacities of all social and request edges are rational numbers. This setting
is of important interest in general, since the parameters of most practical problems are rational numbers. Then problem \eqref{prb1} with rational parameters can be equivalently converted to
one with \emph{integral} parameters by multiplying with a suitably large integer\footnote{For example, it can be the least common multiple of the denominators in the fractional forms of the
rational numbers.} $K$. The solution of the original problem (with rational parameters) is equal to the solution of the new problem (with integral parameters) divided by $K$.

For problem \eqref{prb1} with rational parameters, let $\overline{U}$ and $\overline{R}$ denote the maximum utility and maximum capacity of a request edge, respectively (i.e., $\overline{U} =
\max_{e_{ij}\in E^R} U_{ij}$, $\overline{R} = \max_{e_{ij}\in E^R} R_{ij}$). The following theorem shows that Algorithm~\ref{al:cycle_canceling} is correct and computationally efficient when the service
divisible.
\begin{thm}\label{thm:optimal}
For problem \eqref{prb1} with divisible sensing service and rational parameters, Algorithm~\ref{al:cycle_canceling} computes the optimal flow and has running time
$O(|V||E^R|(|E^R|+|E^S|)\overline{R}\overline{U}K^2)$.
\end{thm}

\textbf{Proof:} As discussed earlier, we first equivalently convert the problem to one with integral parameters by multiplying them by an integer $K$.

Since the capacities of all edges in the graph are integral and the initial empty flow is integral, the residual capacity of the cycle found in the first iteration of the algorithm is
integral, and hence the flow after augmentation is integral. Thus, by induction, the updated flow after each iteration is also integral. This shows that the algorithm finds an integral flow
when it terminates, which is optimal by Lemma~\ref{lm:optimal}.

The utility of the initial empty flow is 0. The utility of any flow is upper bounded by the utility of the flow that saturates all request edges, which is $|E^R|\overline{R}\overline{U}K^2$.
Since the capacities of all edges are integral, the flow utility increases by an integer no less than one at each iteration of Algorithm~\ref{al:cycle_canceling}. Therefore, it takes the
algorithm at most $|E^R|\overline{R}\overline{U}K^2$ iterations to terminate. Since each iteration has running time $O(|V|(|E^S| + |E^R|)$, the desired result follows. $\hfill\square$

In many practical situations, sensing service is indivisible such that the optimization variables of problem~\eqref{prb1} have to be integers. In this case, we can equivalently convert
problem~\eqref{prb1} with rational parameters to one with integer parameters by rounding the capacities of all social and request edges to their respective nearest integers below (i.e.,
taking the floor function) and multiplying the utilities of all request edges by a suitably large integer $K$. Using a similar proof as that of Theorem~\ref{thm:optimal}, we have the
following result.
\begin{thm}\label{thm:optimal_indivisible}
For problem \eqref{prb1} with indivisible sensing service and rational parameters, Algorithm~\ref{al:cycle_canceling} computes the optimal flow and has running time
$O(|V||E^R|(|E^R|+|E^S|)\lfloor \overline{R} \rfloor\overline{U}K)$.
\end{thm}

In Section \ref{sc:simulation}, simulation results demonstrate that the running time of Algorithm~\ref{al:cycle_canceling} is much lower than the above bound.

\textbf{Remark \remark:} The underlying rational of Algorithm~\ref{al:cycle_canceling} can be interpreted as follows. In each iteration of Algorithm~\ref{al:cycle_canceling}, the flow in the
social-request graph is augmented with a balanced flow along a cycle of positive weight in the residual graph. For each edge with positive weight in that cycle, the utility of flow on the
corresponding request edge increases, while for each edge with negative weight in that cycle, that utility decreases. Since the total weight of the edges in the cycle is positive, the total
utility of flow increases. In other words, a balanced flow along the positive weight cycle captures \emph{the tradeoff between increasing the utilities on some request edges and decreasing
the utilities on some other request edges such that the total utility increases}. Note that although the flows on social edges does not generate utility, the edges with zero weights in the
residual graph, which are constructed from the social edges, contribute to forming a cycle, and hence the utility obtained on request edges.

\textbf{Remark \remark:} It is worth noting that, when sensing service is indivisible, problem~\eqref{prb1} is essentially an integer linear program (ILP), which is NP-hard to solve in general.
However, using a network flow approach, we can capture and exploit the specific combinatorial structure of the problem, based on which a polynomial-time algorithm can be developed to solve~it.

\section{Performance Evaluation}\label{sc:simulation}

In this section, we provide simulation results to evaluate the performance of the STAR mechanism. We compare STAR with two incentive mechanisms as benchmarks, which use social trust only and
reciprocity only, respectively:
\begin{itemize}

  \item \emph{Social trust based mechanism} (ST): Under this mechanism, a user is willing to provide service to another if and only if it receives social credit from that user or an
      intermediate user. Therefore, in the social-request graph, this mechanism can use a cycle consisting of social edges and \emph{exactly one} request edge (e.g., as illustrated in
      Fig.~\ref{fg:cooperation_cycle}(c),(d));

  \item \emph{Reciprocity based mechanism} (RP): Under this mechanism, a user is willing to provide service if and only if it also receives service from another. Therefore, in the
      social-request graph, this mechanism can use a cycle consisting of \emph{only} request edges  (e.g., as illustrated in Fig.~\ref{fg:cooperation_cycle}(a),(b)).
\end{itemize}
We observe that each benchmark mechanism only uses a \emph{subset} of the incentive structures (i.e., the cycles in the social-request graph) used in the STAR mechanism. Note that the
incentive mechanism only using social trust is equivalent to that in~\cite{Liu10}.

\begin{figure}[t]
	\centering
        \subfigure[ ]{\hspace{-0.4em}\includegraphics[scale=0.30]{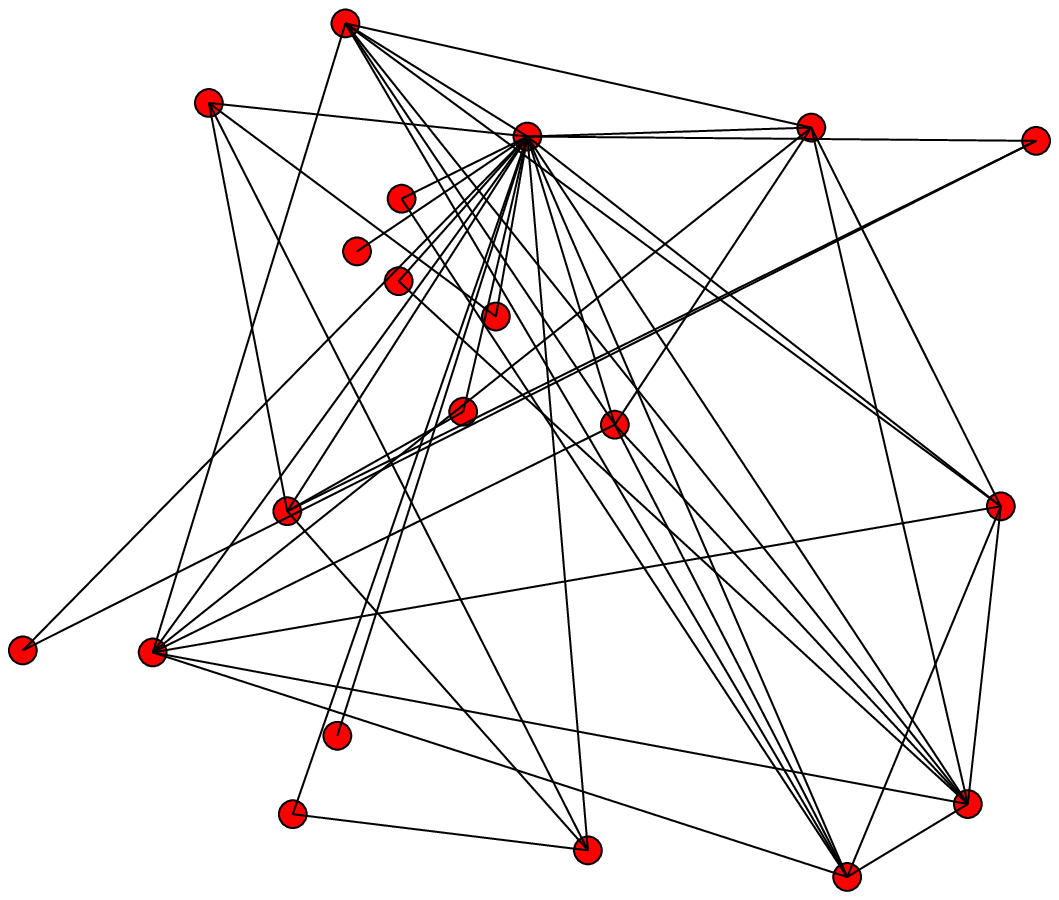}\hspace{-0.8em}}%
        \subfigure[ ]{\includegraphics[scale=0.30]{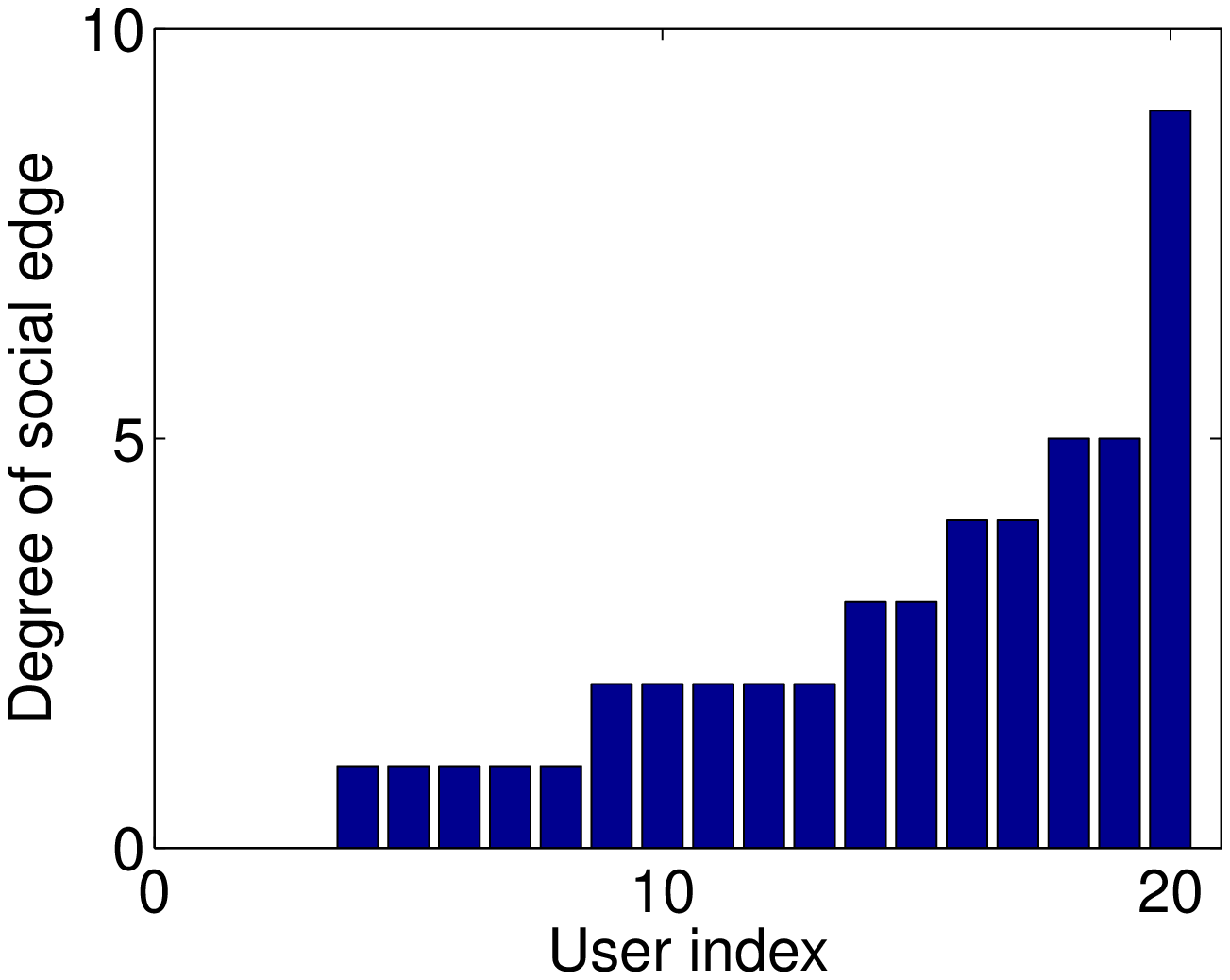}}%
        \caption{A social network of 20 users in real dataset Brightkite \cite{Brightkite}: (a) social network structure; (b) degree of social edge.} \label{fg:real_social_graph}
        \vspace{-0.5em}
\end{figure}


\begin{figure}[t]
	\centering
        \subfigure[ ]{\hspace{-0.4em}\includegraphics[scale=0.30]{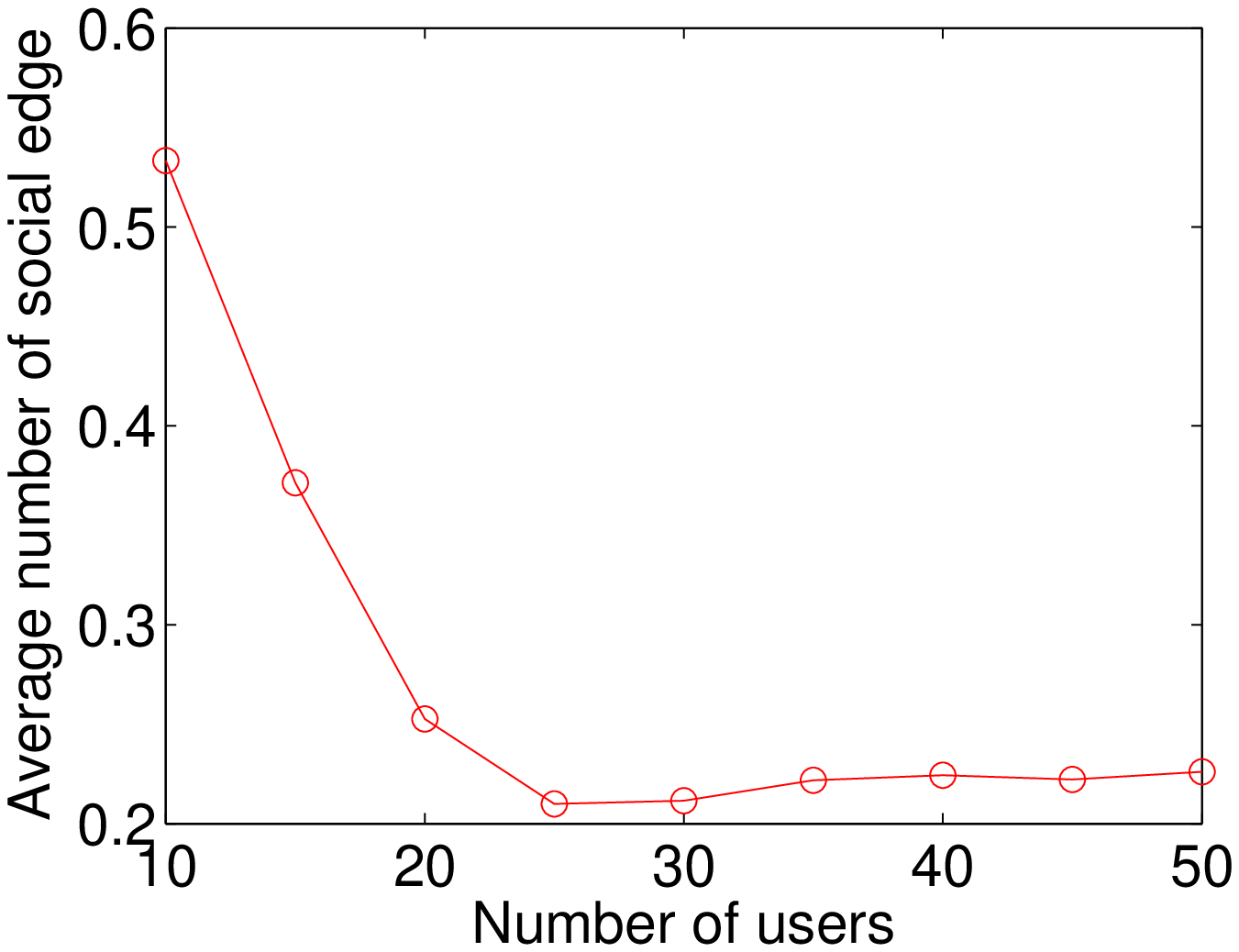}\hspace{-0.8em}}%
        \subfigure[ ]{\includegraphics[scale=0.30]{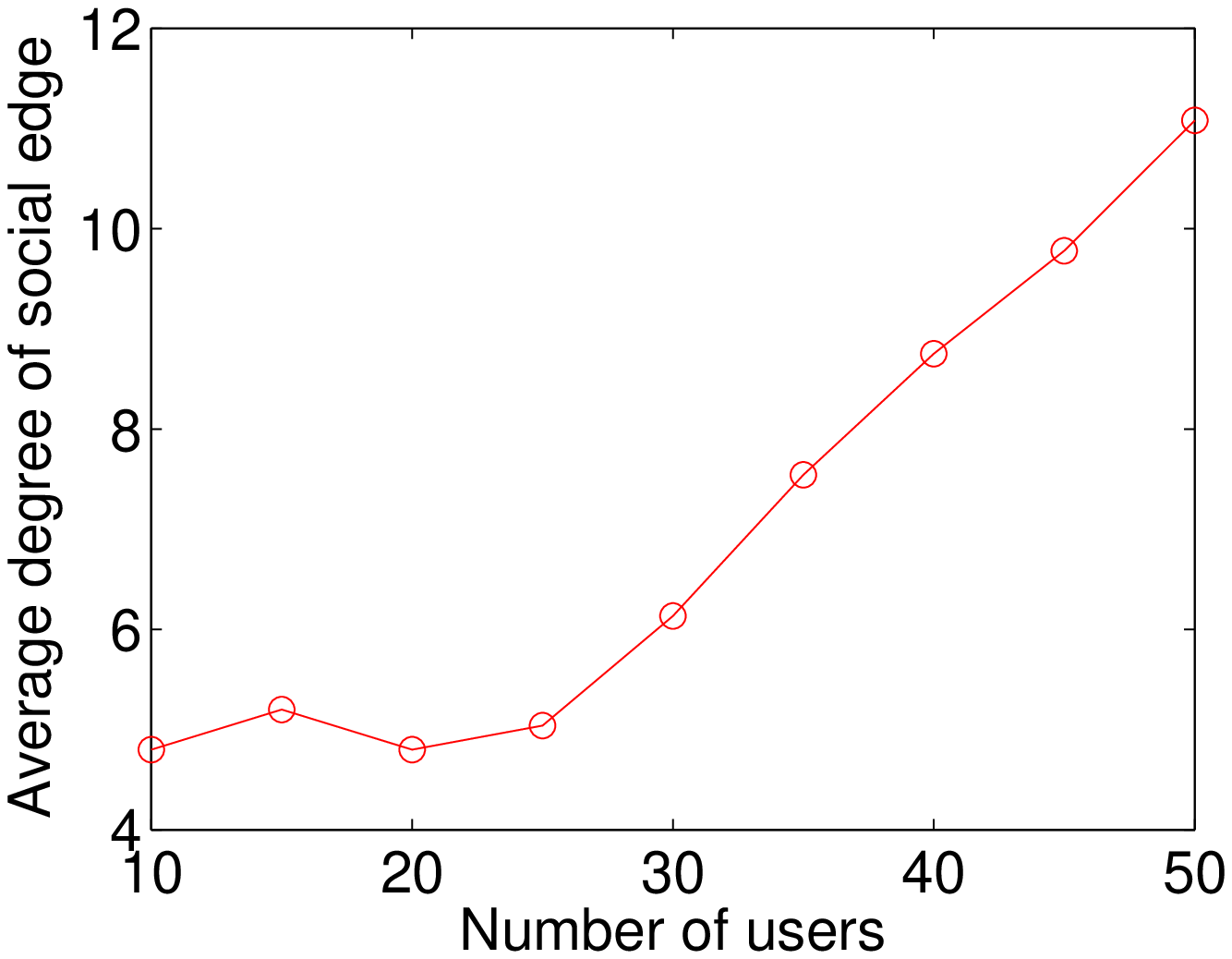}}%
        \caption{(a) Average number of social edge (b) average degree of social edge vs. number of users in real dataset Brightkite \cite{Brightkite}.} \label{fg:real_Ps_N}
        \vspace{-0.5em}
\end{figure}


\begin{figure*}[t]
\begin{minipage}{.33\textwidth}
\includegraphics[width=1\textwidth]{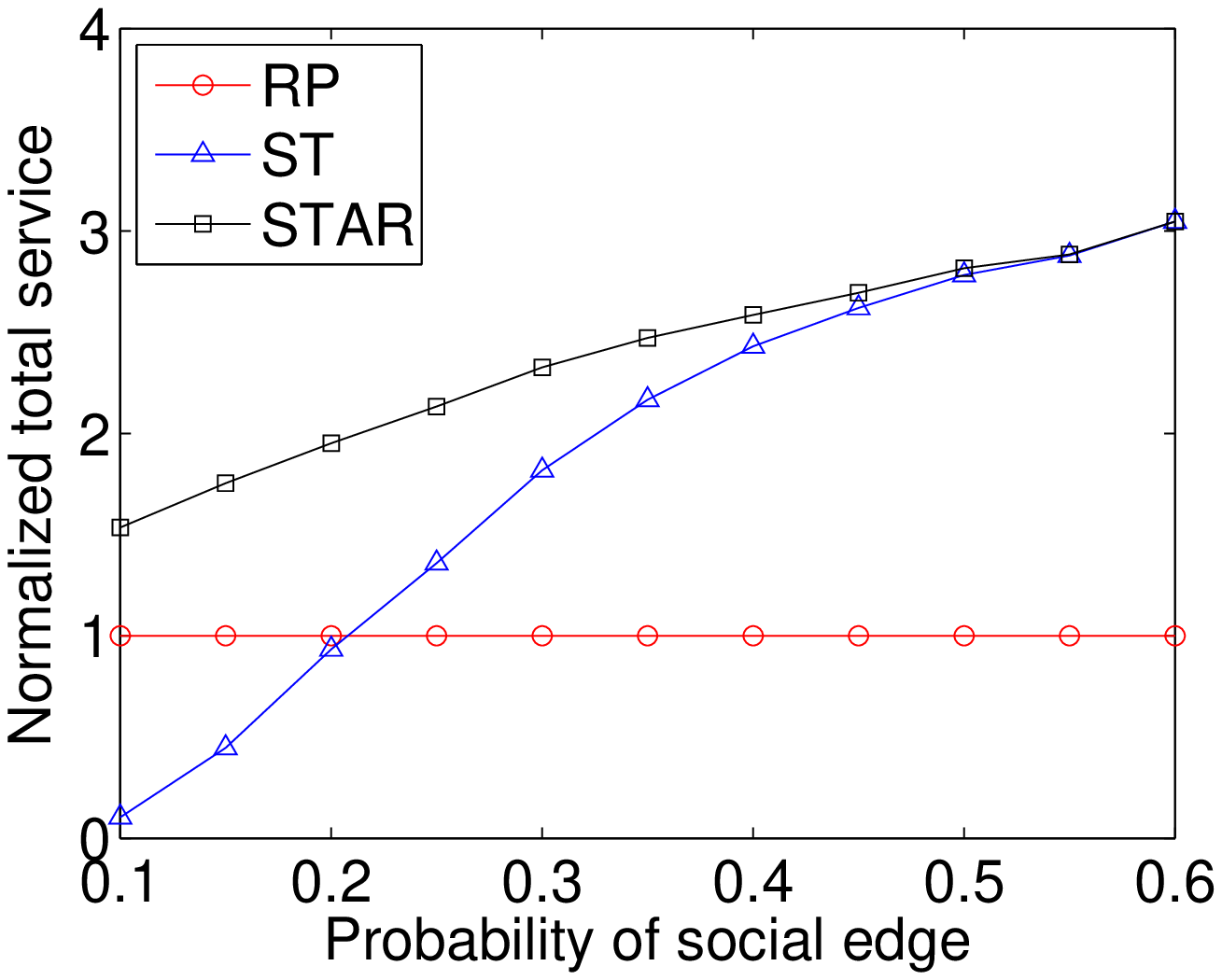}
\caption{Impact of $P_S$ on total service amount for the random setting.}
\label{fg:total_service_Ps}
\end{minipage}
\begin{minipage}{.33\textwidth}
\includegraphics[width=1\textwidth]{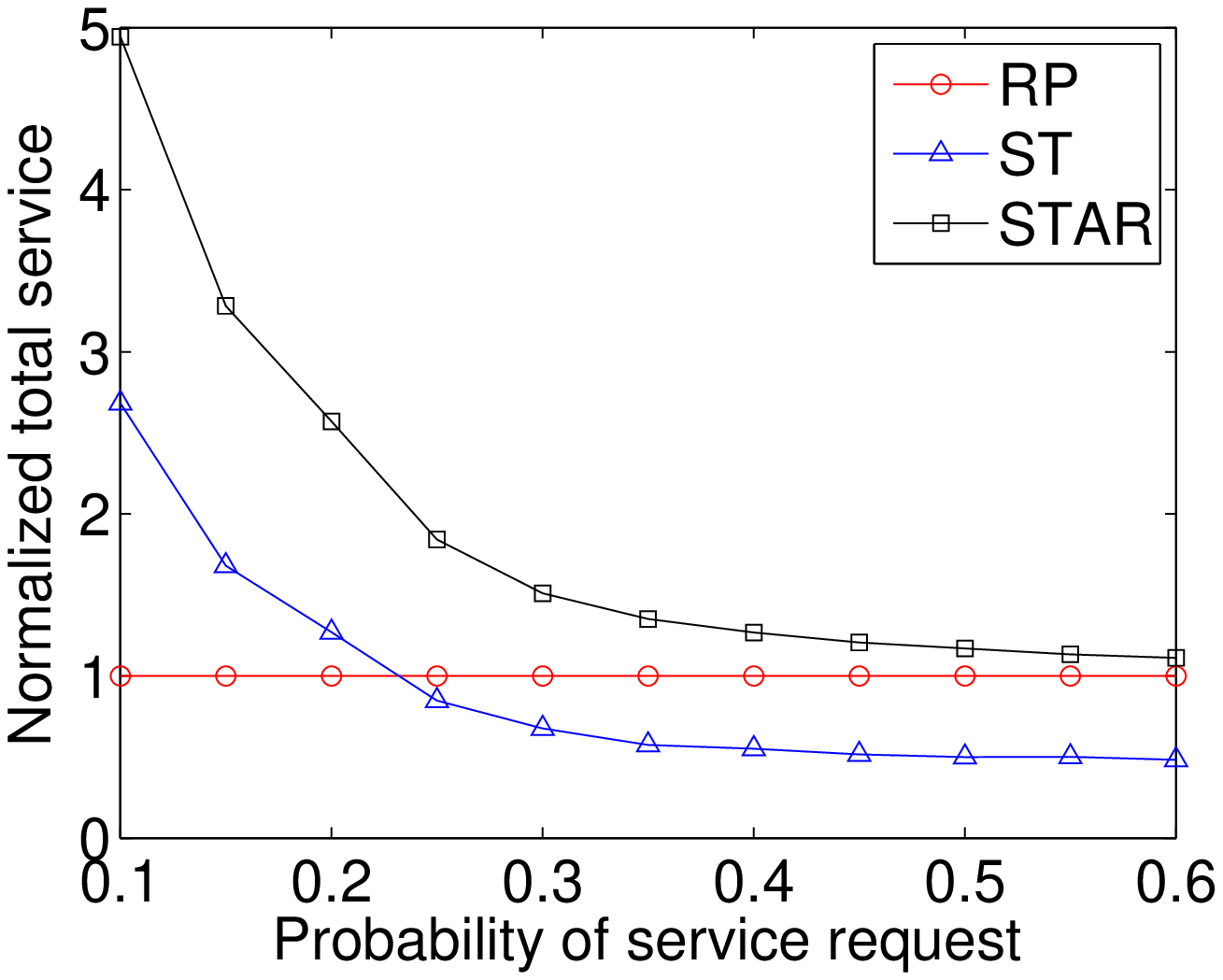}
\caption{Impact of $P_R$ on total service amount for the random setting.}
\label{fg:total_service_Pd}
\end{minipage}
\begin{minipage}{.33\textwidth}
\includegraphics[width=1\textwidth]{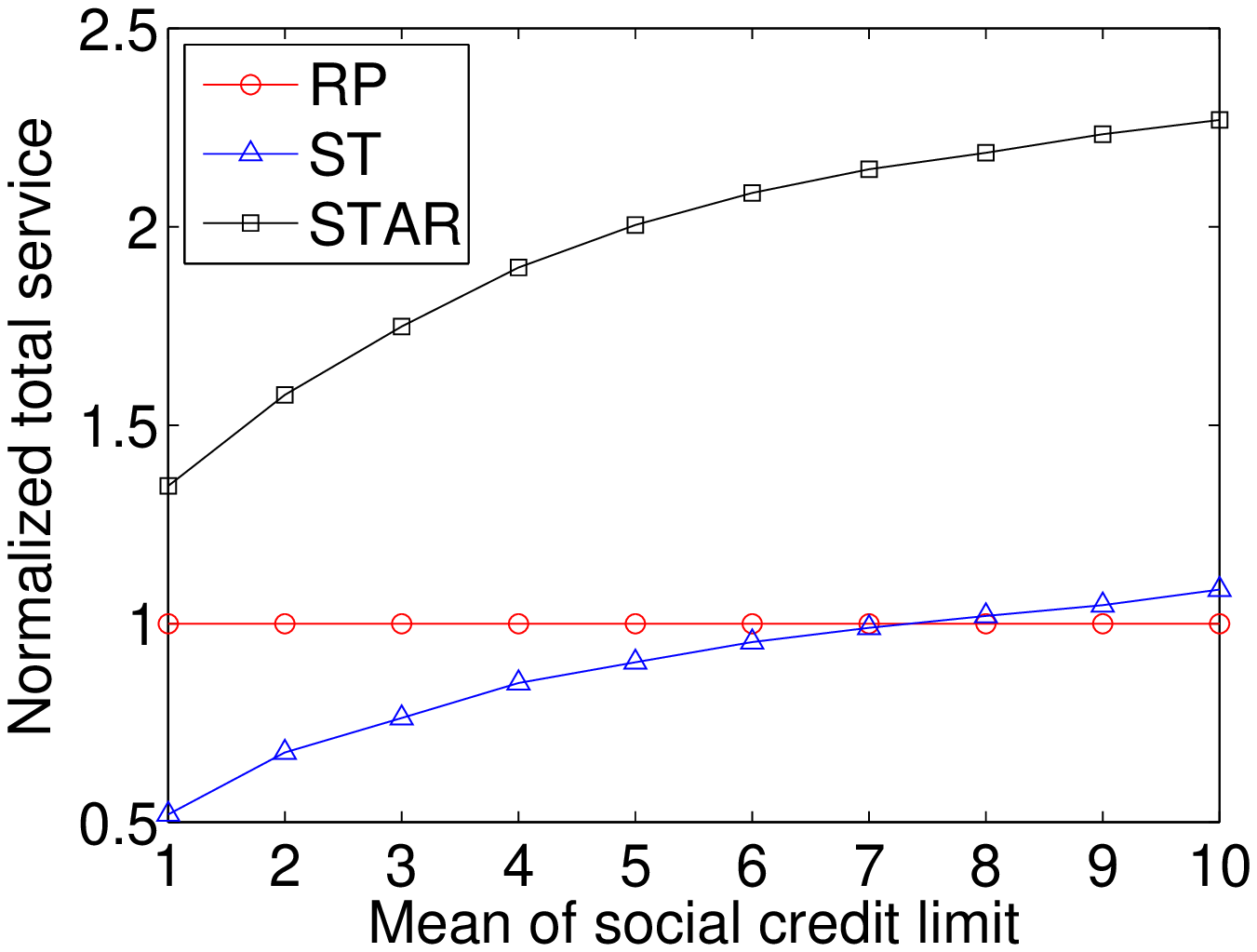}
\caption{Impact of $\mu_S$ on total service amount for the random setting.}
\label{fg:total_service_Cs}
\end{minipage}
\end{figure*}

\begin{figure*}[t]
\begin{minipage}{.33\textwidth}
\includegraphics[width=1\textwidth]{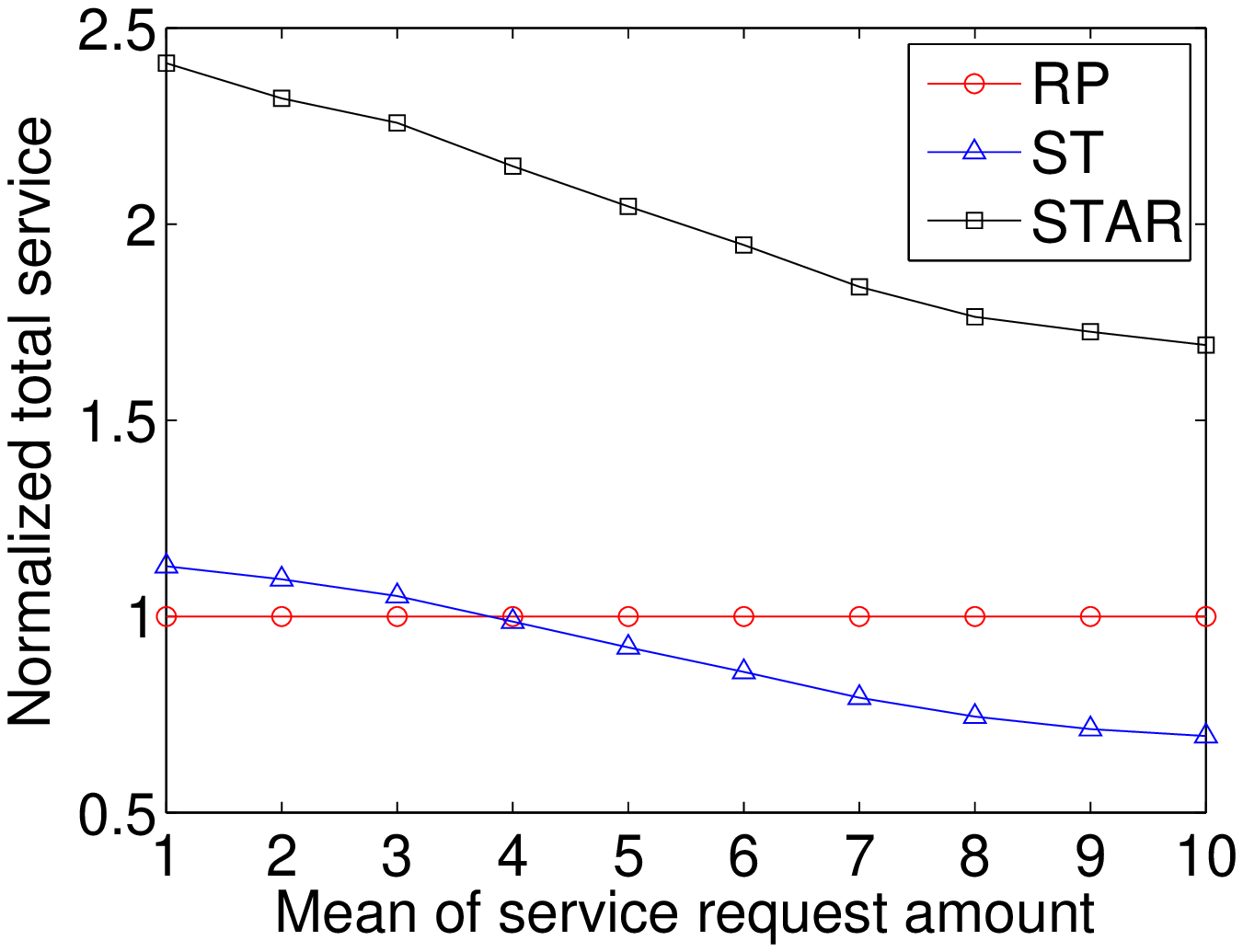}
\caption{Impact of $\mu_R$ on total service amount for the random setting.}
\label{fg:total_service_Cd}
\end{minipage}
\begin{minipage}{.33\textwidth}
\includegraphics[width=1\textwidth]{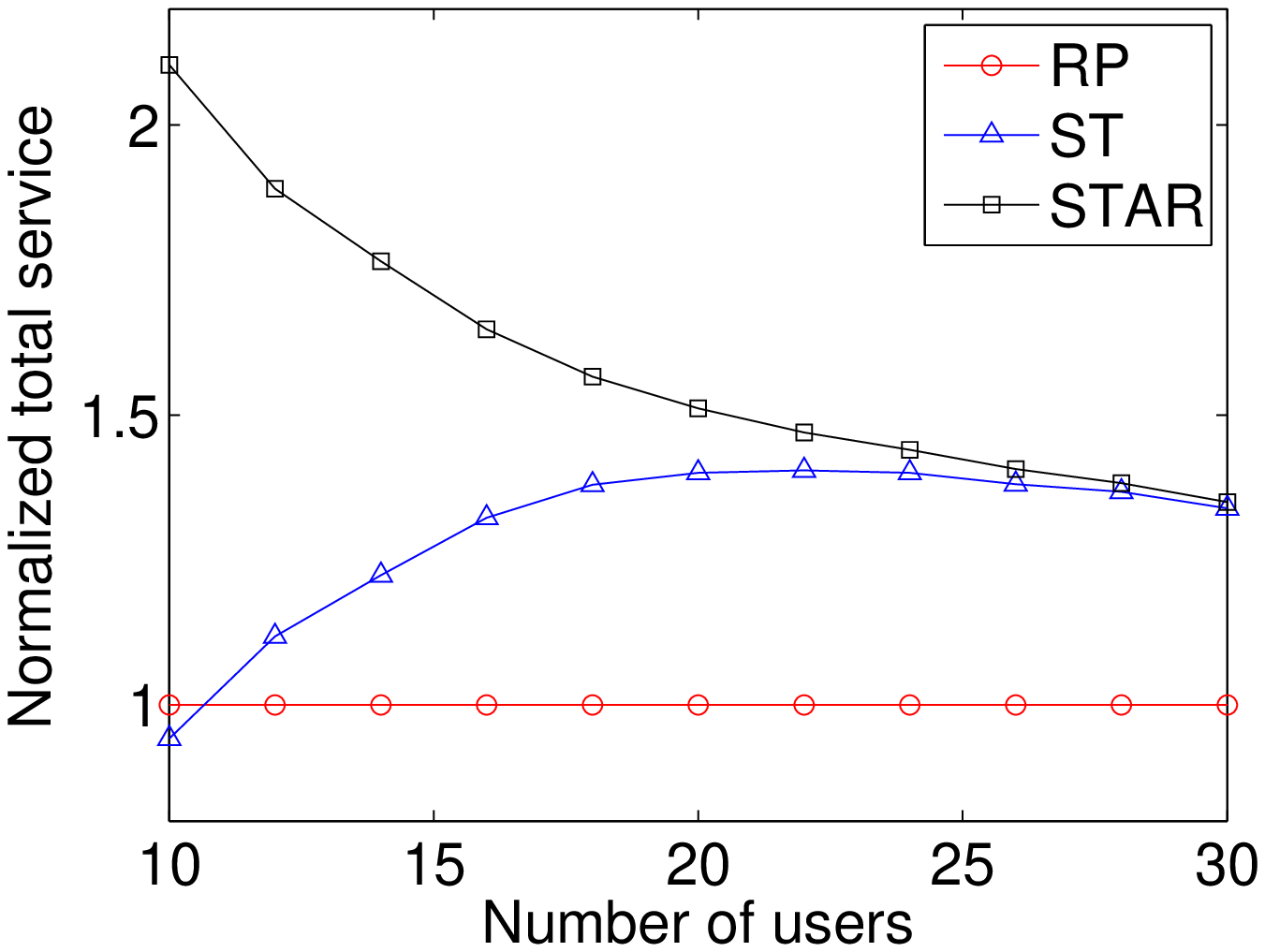}
\caption{Impact of $N$ on total service amount for the random setting.}
\label{fg:total_service_N}
\end{minipage}
\begin{minipage}{.33\textwidth}
\includegraphics[width=1\textwidth]{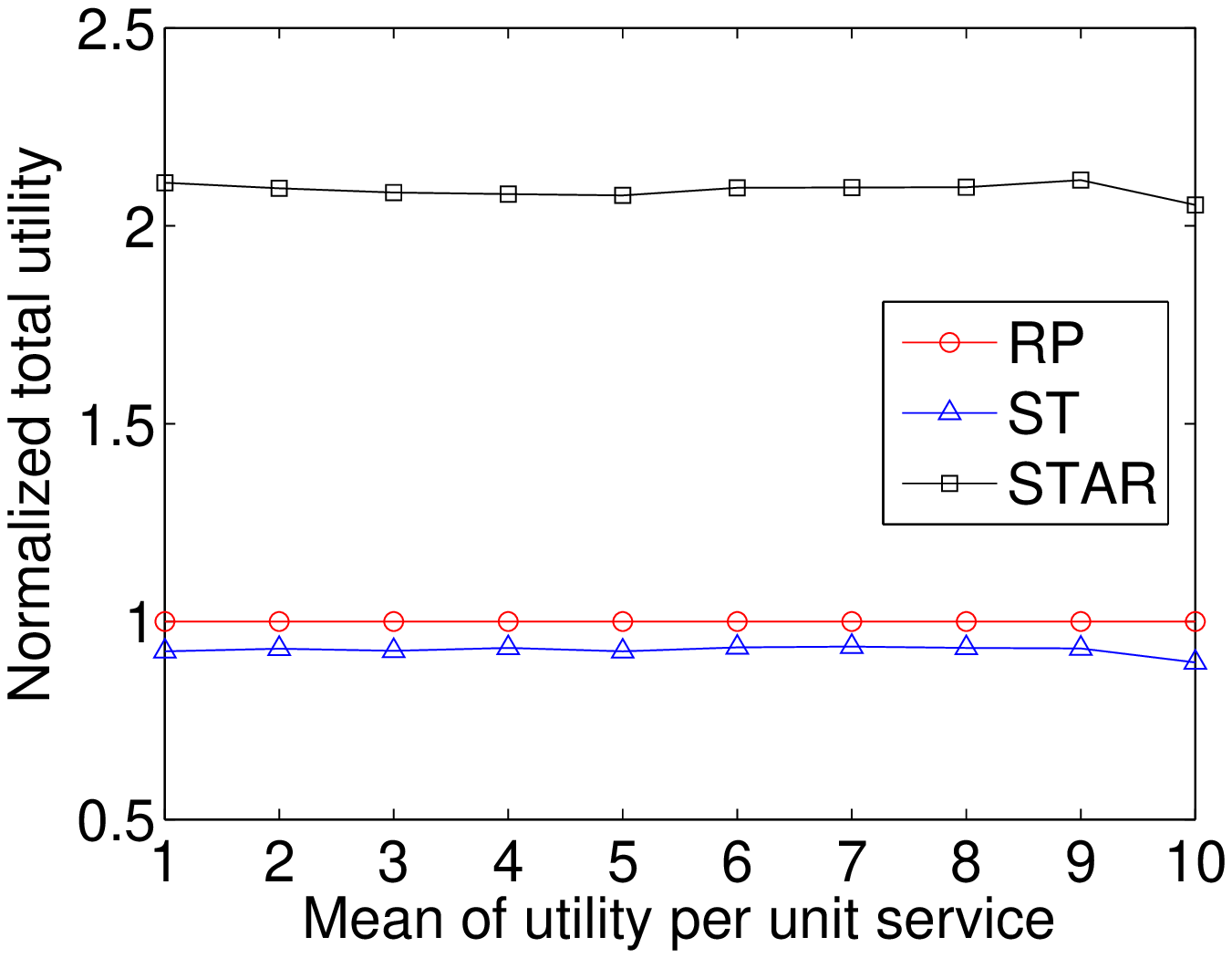}
\caption{Impact of $\mu_U$ on total service utility for the random setting.}
\label{fg:total_utility_Cw}
\end{minipage}
\end{figure*}

\begin{figure*}[t]
\begin{minipage}{.33\textwidth}
\includegraphics[width=1\textwidth]{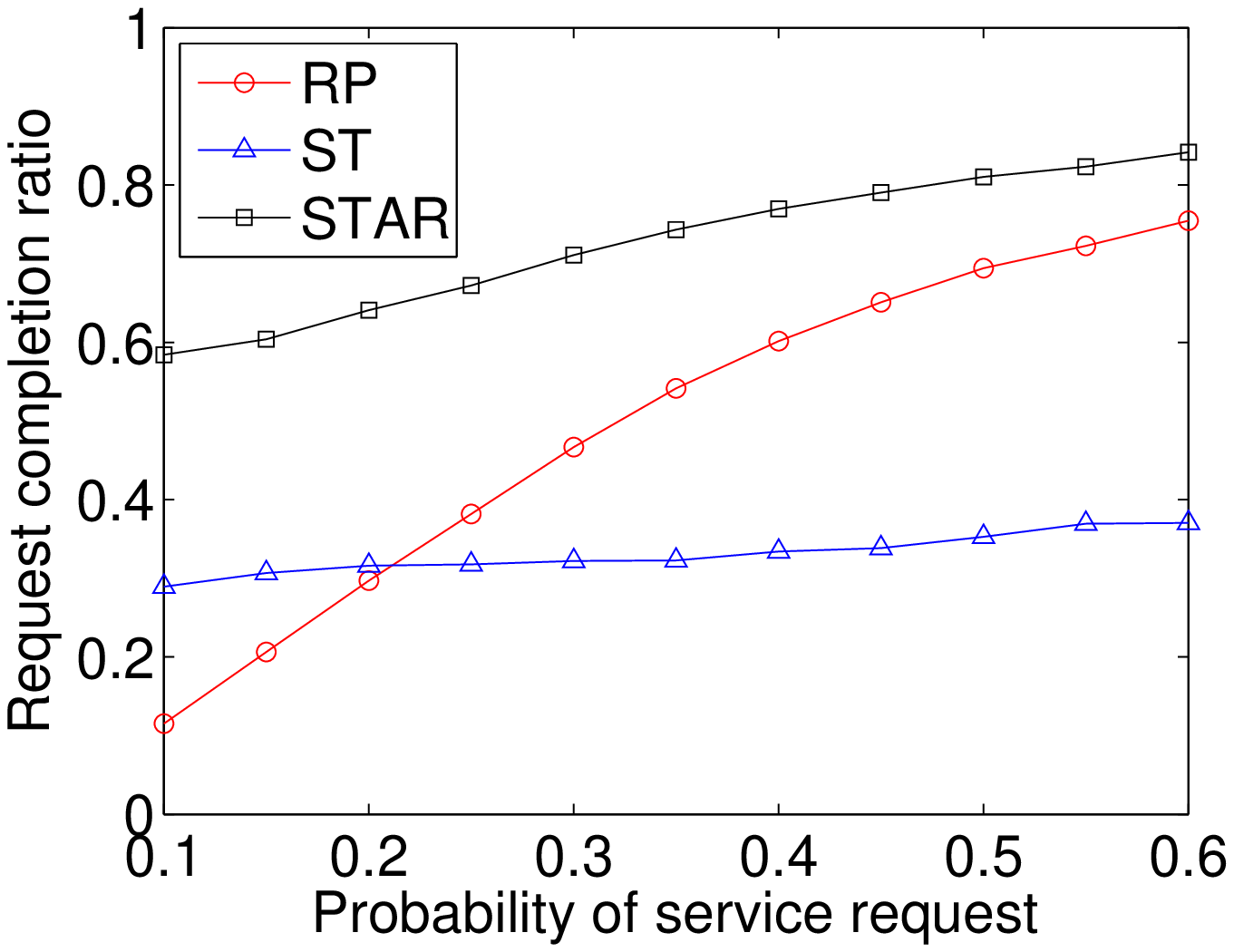}
\caption{Impact of $P_R$ on request completion ratio for the random setting.}
\label{fg:total_ratio_Pd}
\end{minipage}
\begin{minipage}{.33\textwidth}
\includegraphics[width=1\textwidth]{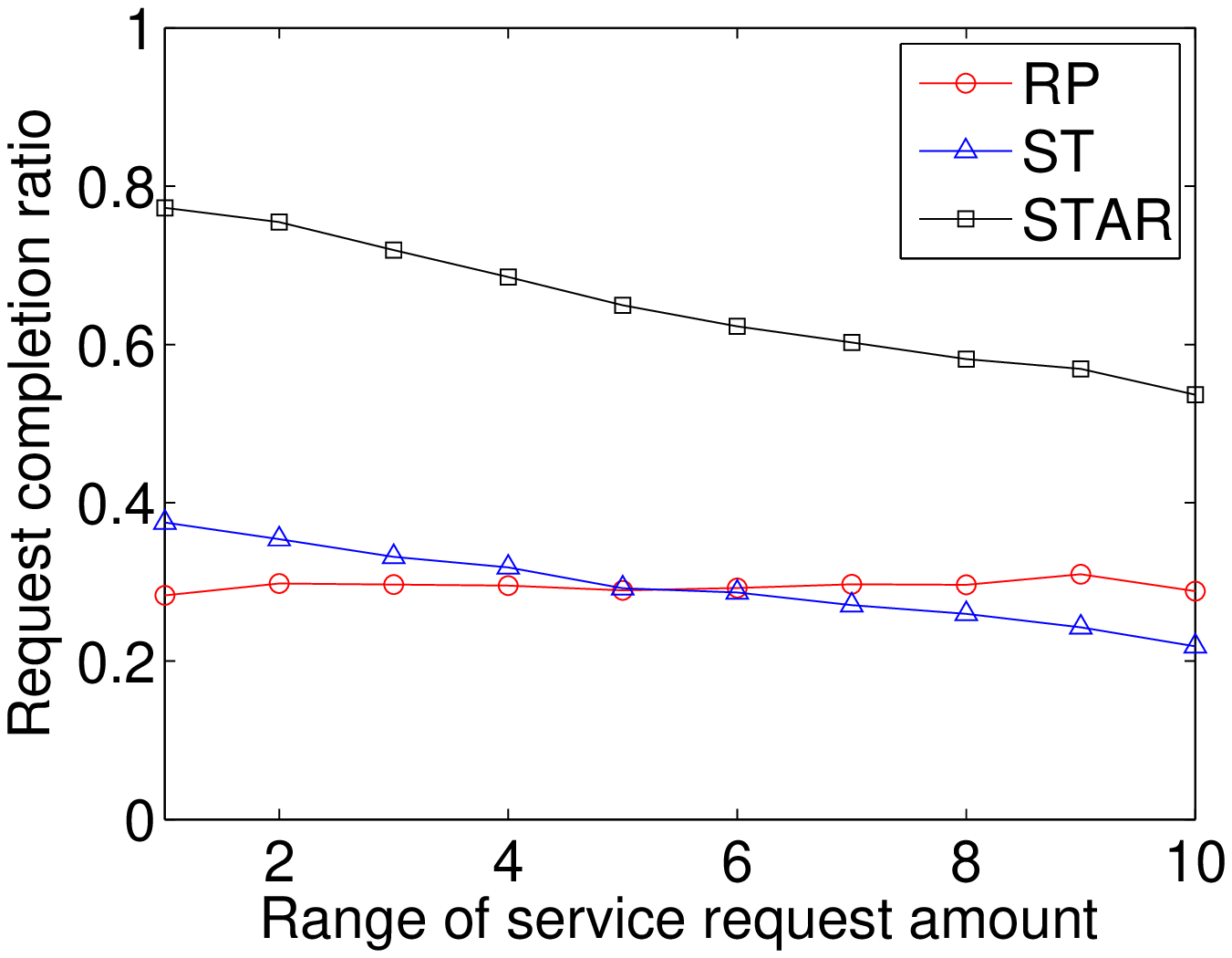}
\caption{Impact of $\mu_R$ on request completion ratio for the random setting.}
\label{fg:total_ratio_Cd}
\end{minipage}
\begin{minipage}{.33\textwidth}
\includegraphics[width=1\textwidth]{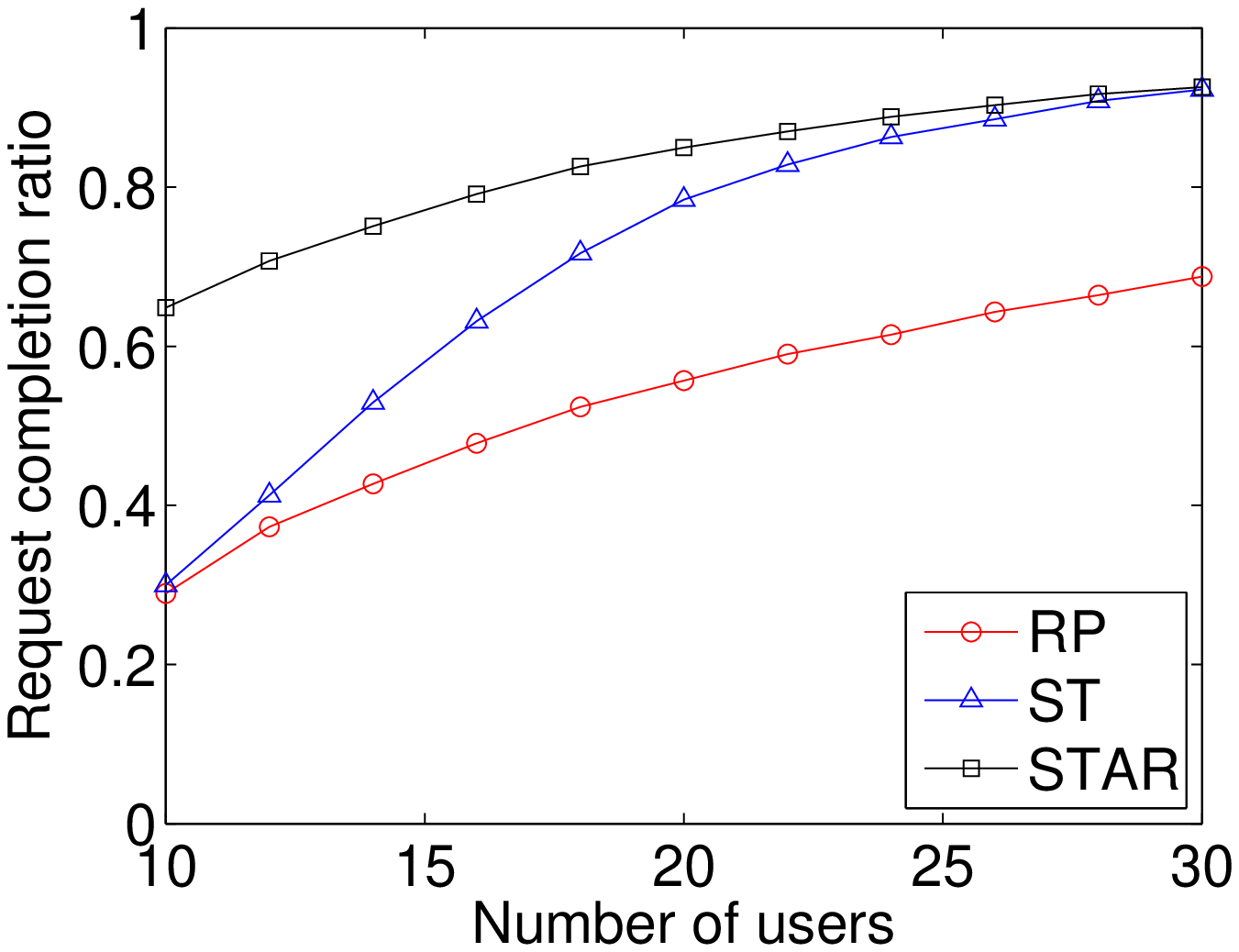}
\caption{Impact of $N$ on request completion ratio for the random setting.}
\label{fg:total_ratio_N}
\end{minipage}
\end{figure*}

\begin{figure*}[t]
\begin{minipage}{.33\textwidth}
\includegraphics[width=1\textwidth]{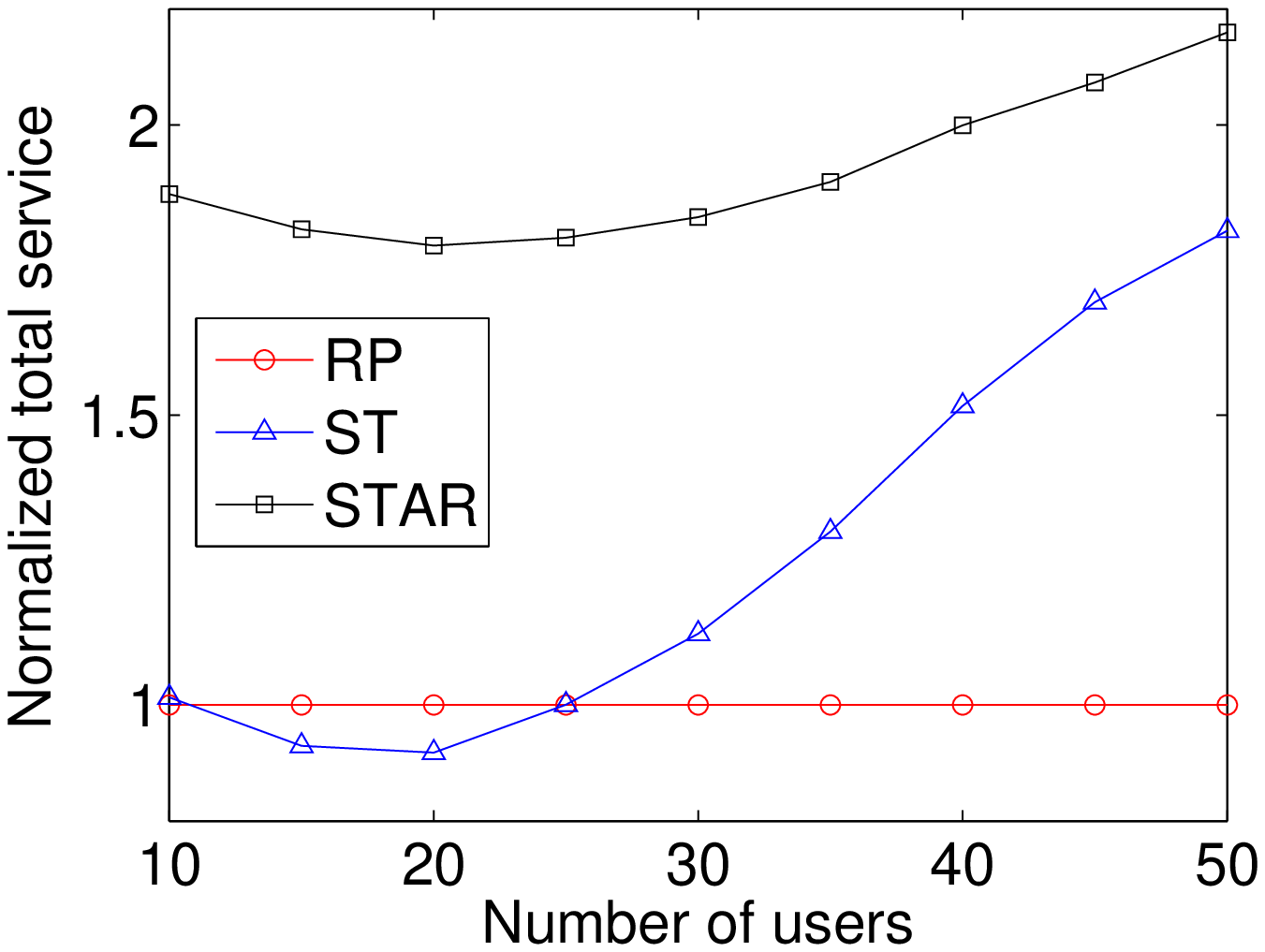}
\caption{Impact of $N$ on total service amount for the practical setting.}
\label{fg:real_service_N}
\end{minipage}
\begin{minipage}{.33\textwidth}
\includegraphics[width=1\textwidth]{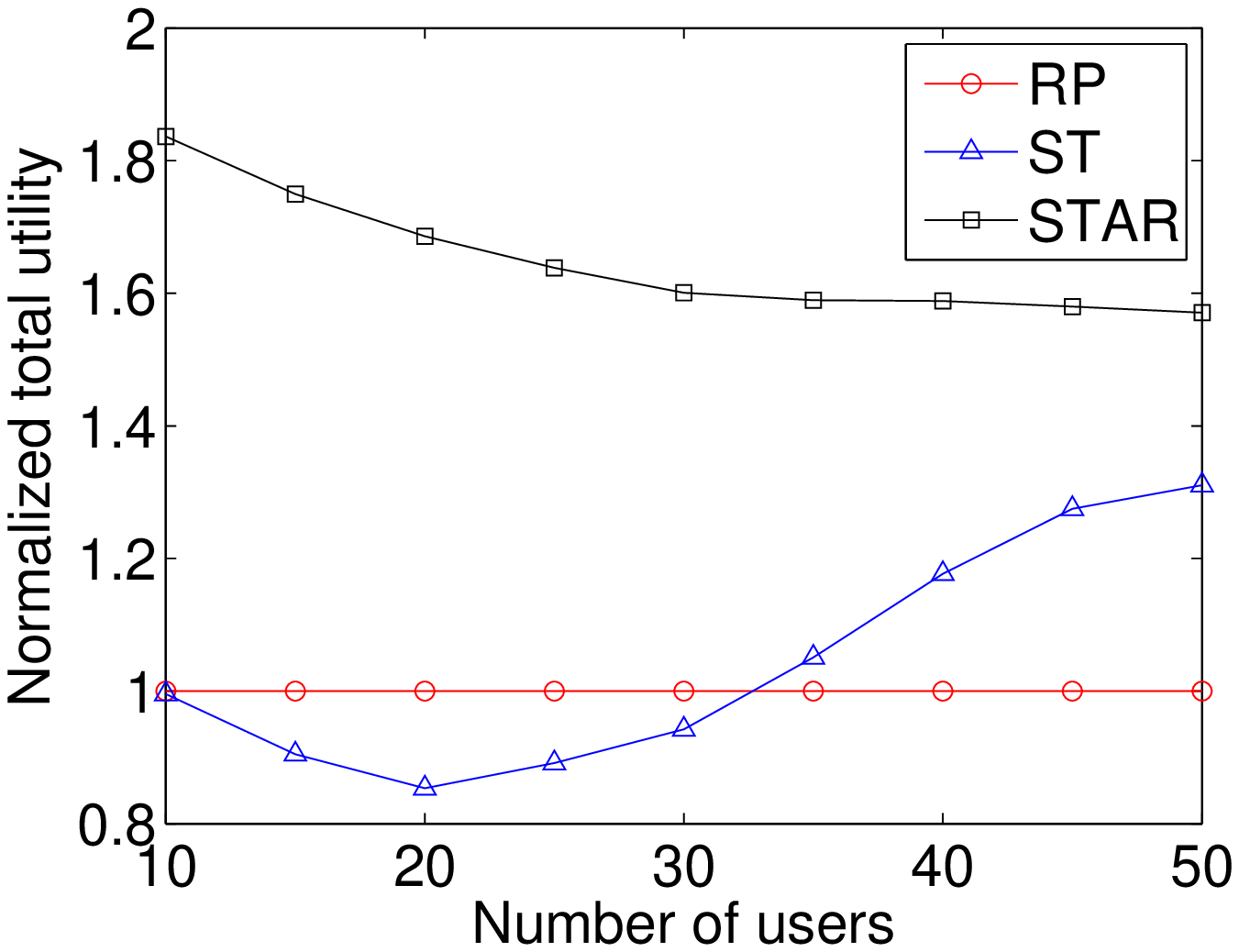}
\caption{Impact of $N$ on total service utility for the practical setting.}
\label{fg:real_utility_N}
\end{minipage}
\begin{minipage}{.33\textwidth}
\includegraphics[width=1\textwidth]{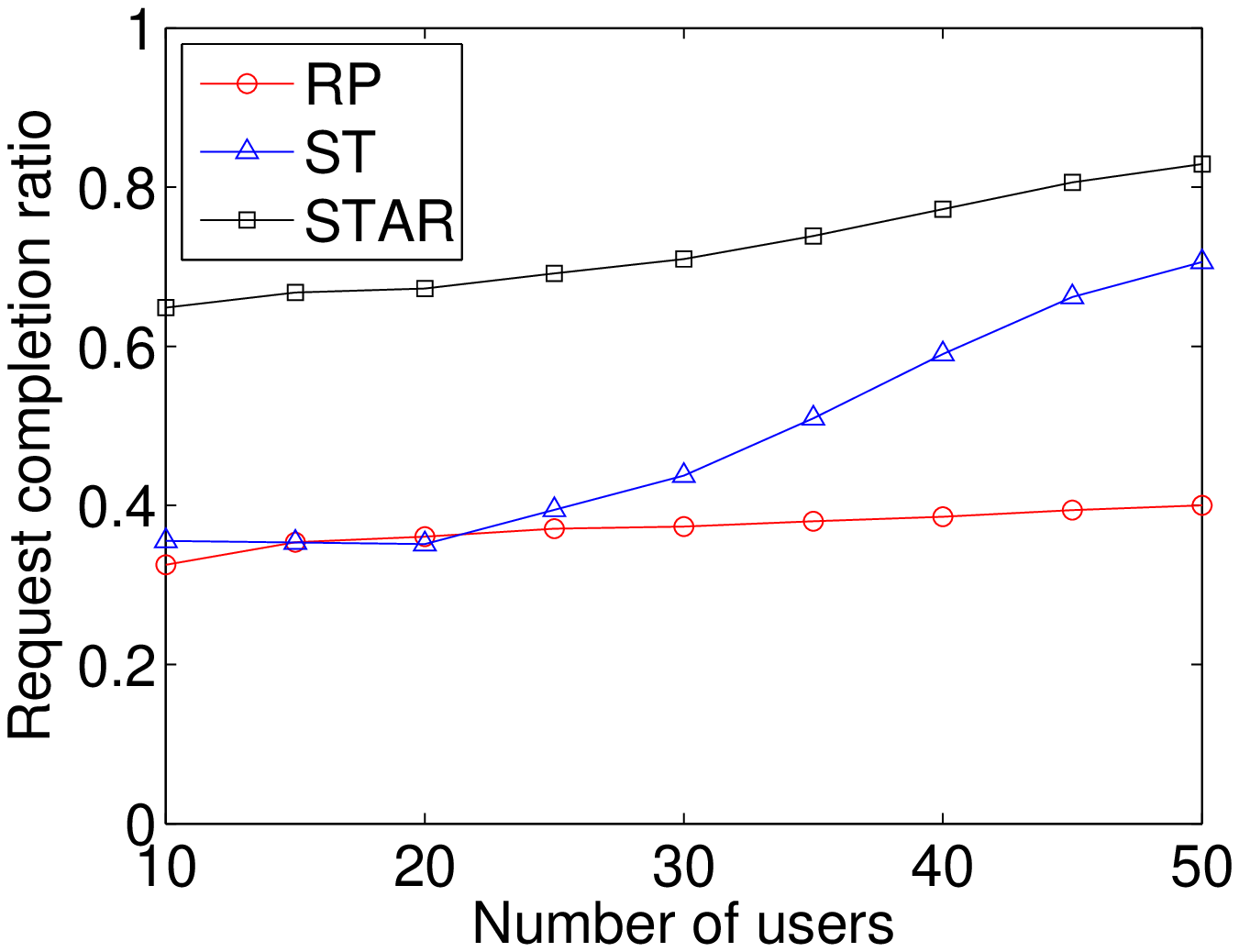}
\caption{Impact of $N$ on request completion ratio for the practical setting.}
\label{fg:real_ratio_N}
\end{minipage}
\end{figure*}

\begin{figure*}[t]
\begin{minipage}{.33\textwidth}
\includegraphics[width=1\textwidth]{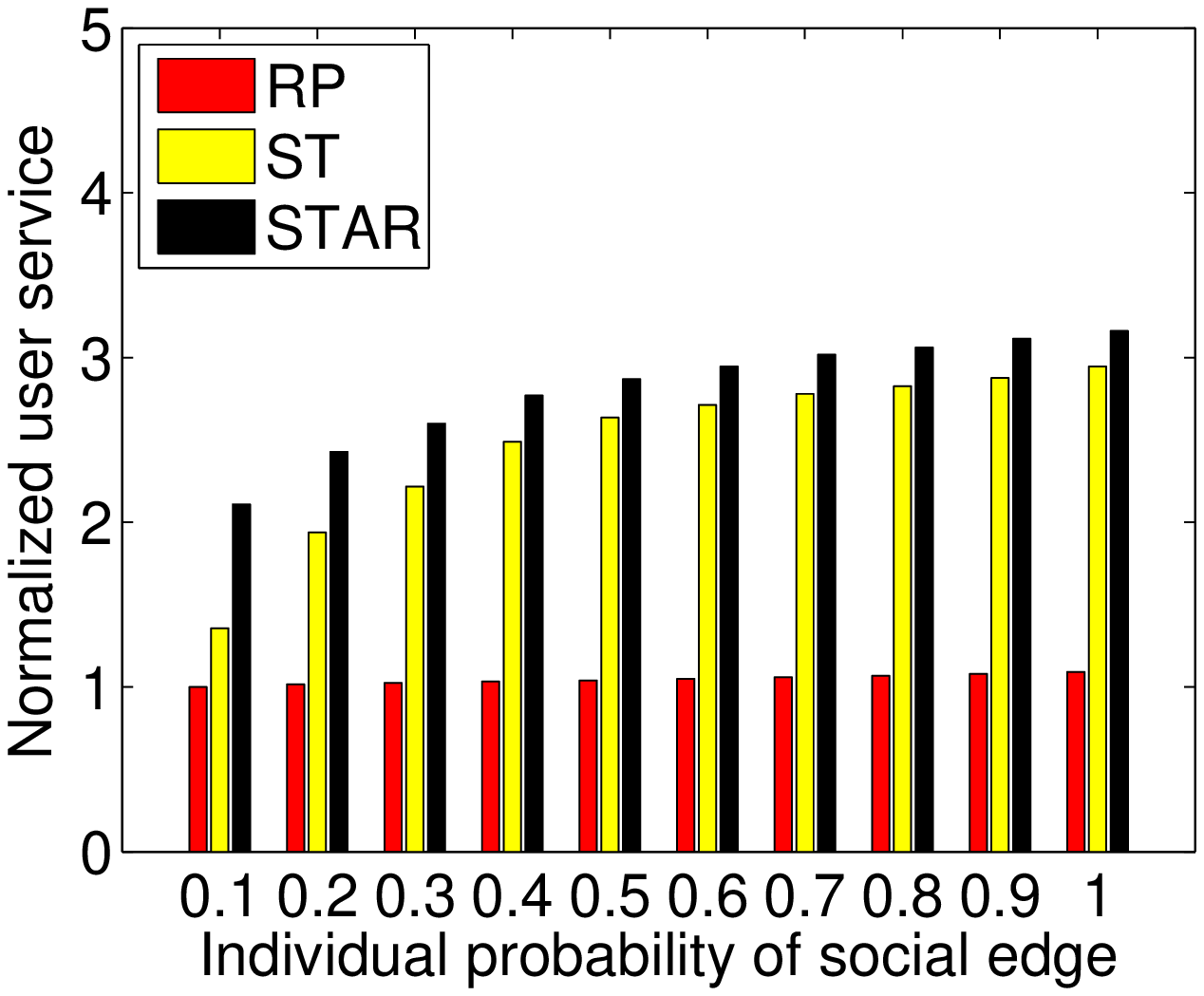}
\caption{Impact of $P_{S_i}$ on user service amount for the random setting.}
\label{fg:user_service_Ps}
\end{minipage}
\begin{minipage}{.33\textwidth}
\includegraphics[width=1\textwidth]{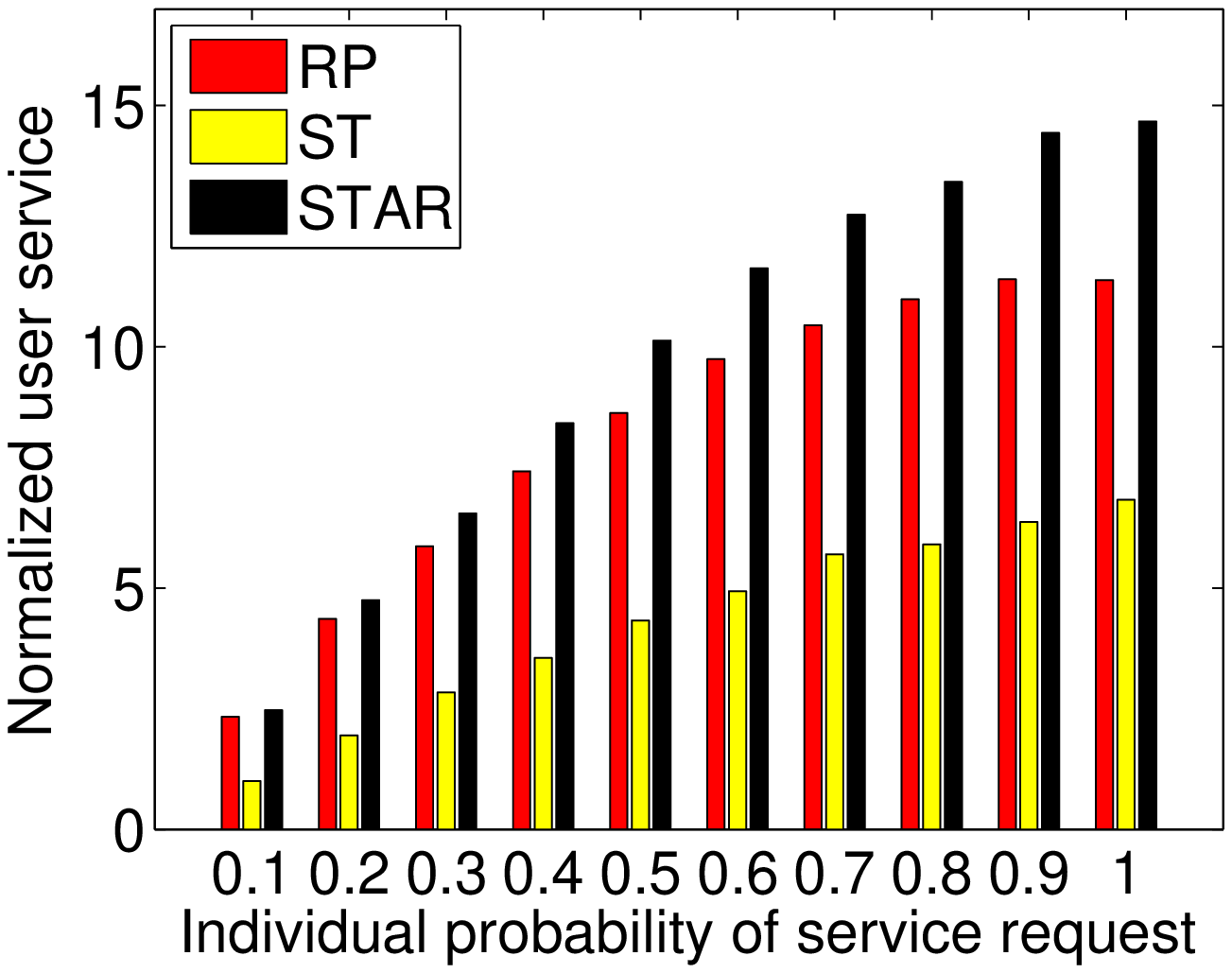}
\caption{Impact of $P_{R_i}$ on user service amount for the random setting.}
\label{fg:user_service_Pd}
\end{minipage}
\begin{minipage}{.33\textwidth}
\includegraphics[width=1\textwidth]{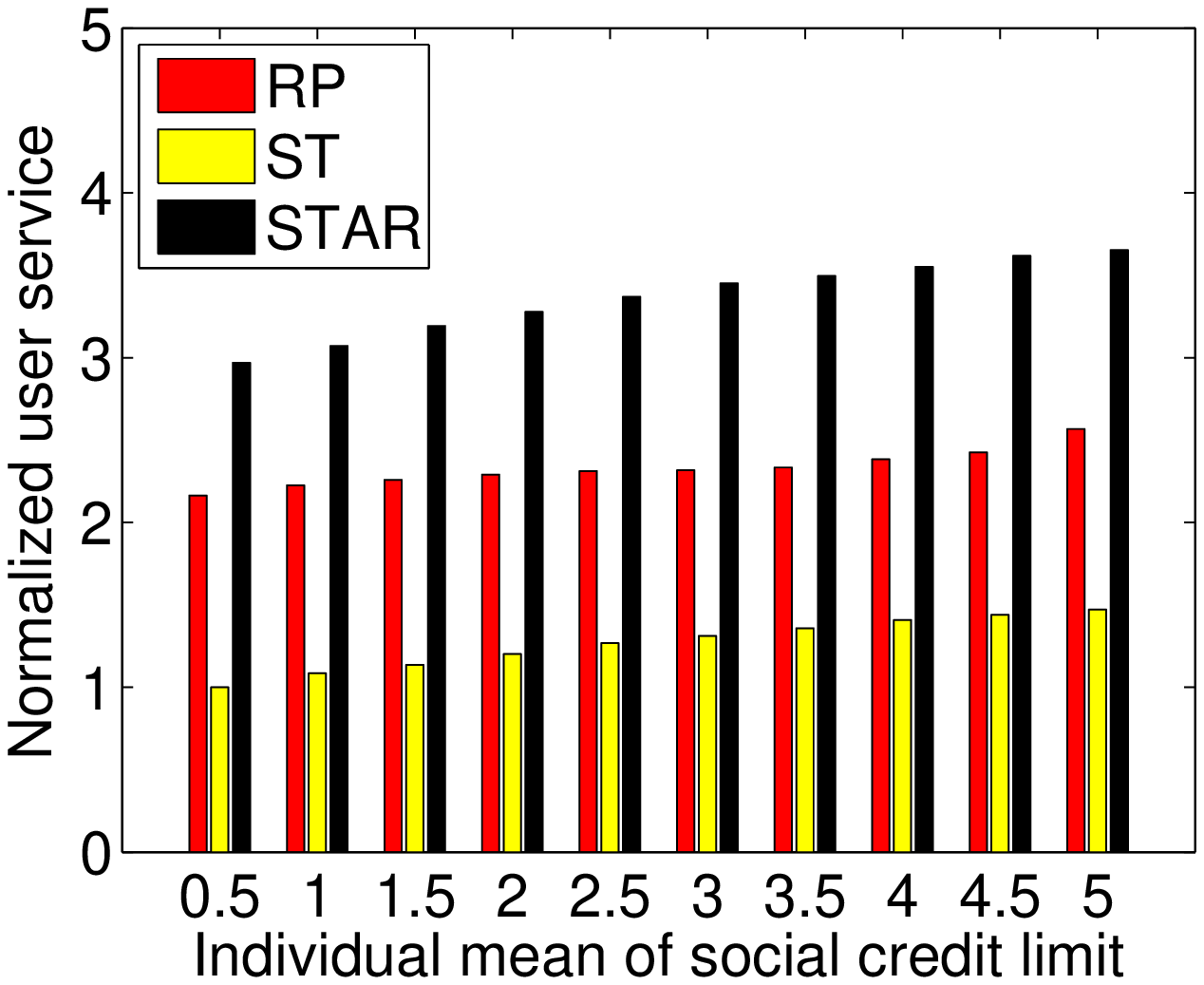}
\caption{Impact of $\mu_{S_i}$ on user service amount for the random setting.}
\label{fg:user_service_Cs}
\end{minipage}
\end{figure*}

\begin{figure*}[t]
\begin{minipage}{.33\textwidth}
\includegraphics[width=1\textwidth]{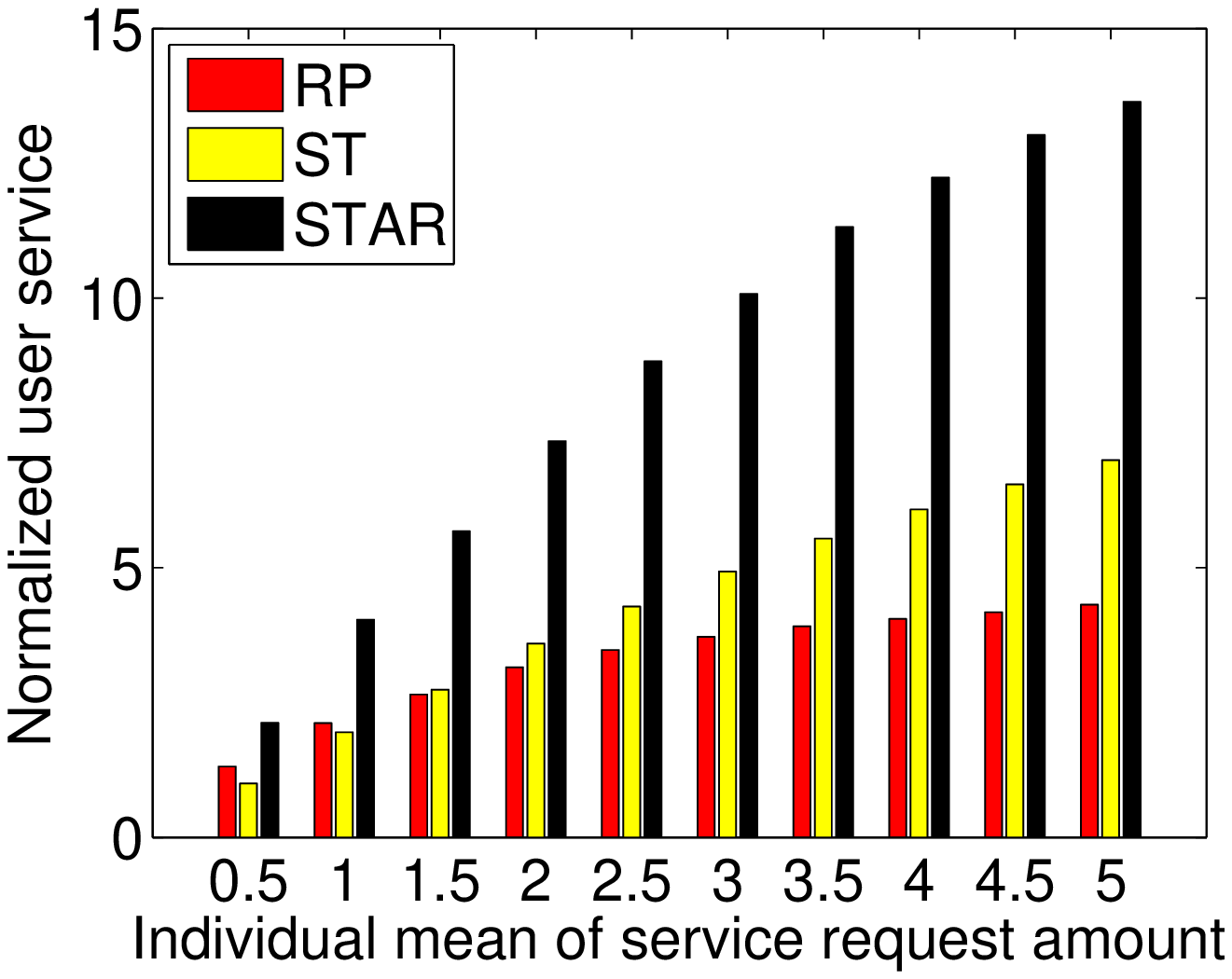}
\caption{Impact of $\mu_{R_i}$ on user service amount for the random setting.}
\label{fg:user_service_Cd}
\end{minipage}
\begin{minipage}{.33\textwidth}
\includegraphics[width=1\textwidth]{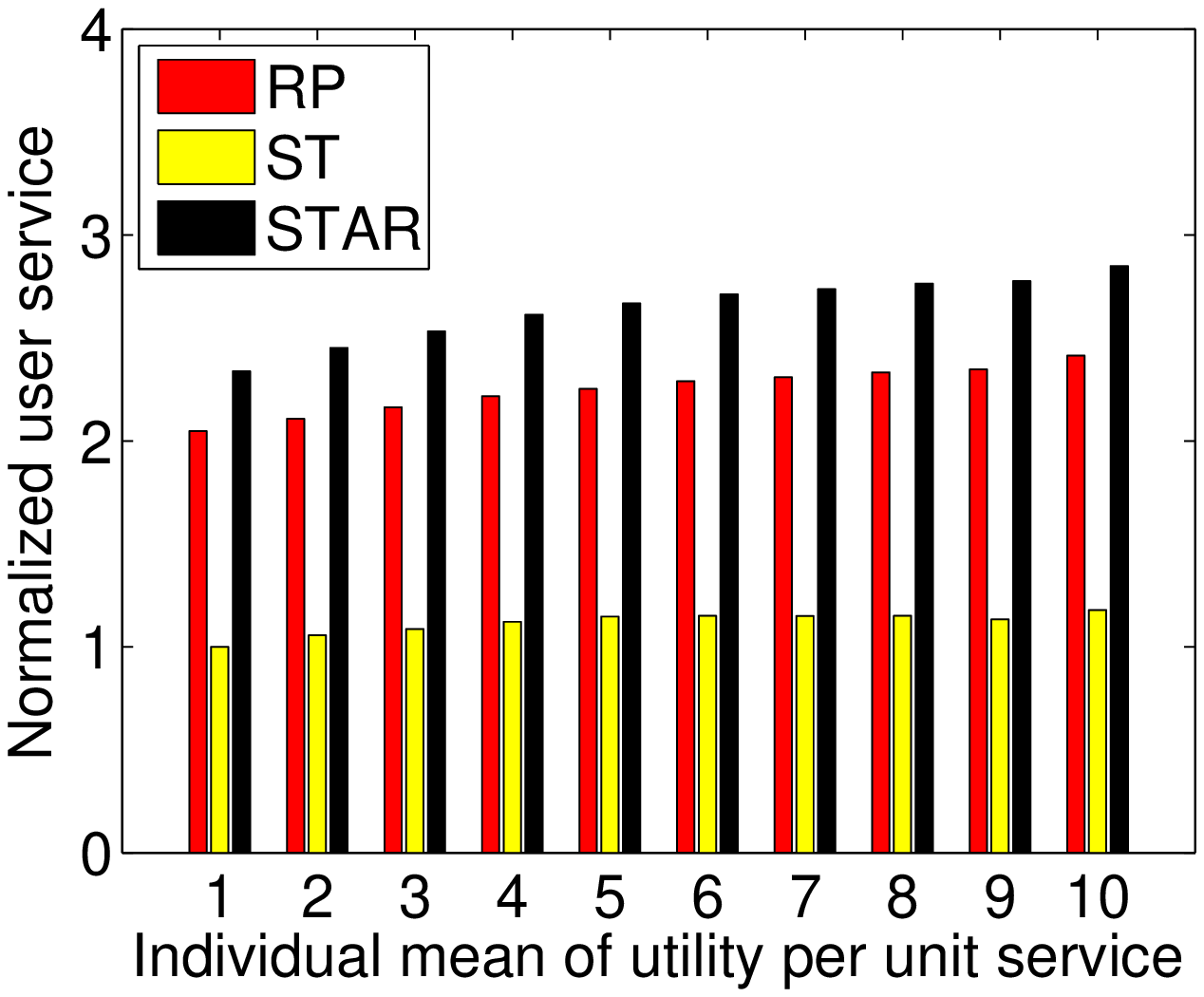}
\caption{Impact of $\mu_{U_i}$ on user service amount for the random setting.}
\label{fg:user_service_Cw}
\end{minipage}
\begin{minipage}{.33\textwidth}
\includegraphics[width=1\textwidth]{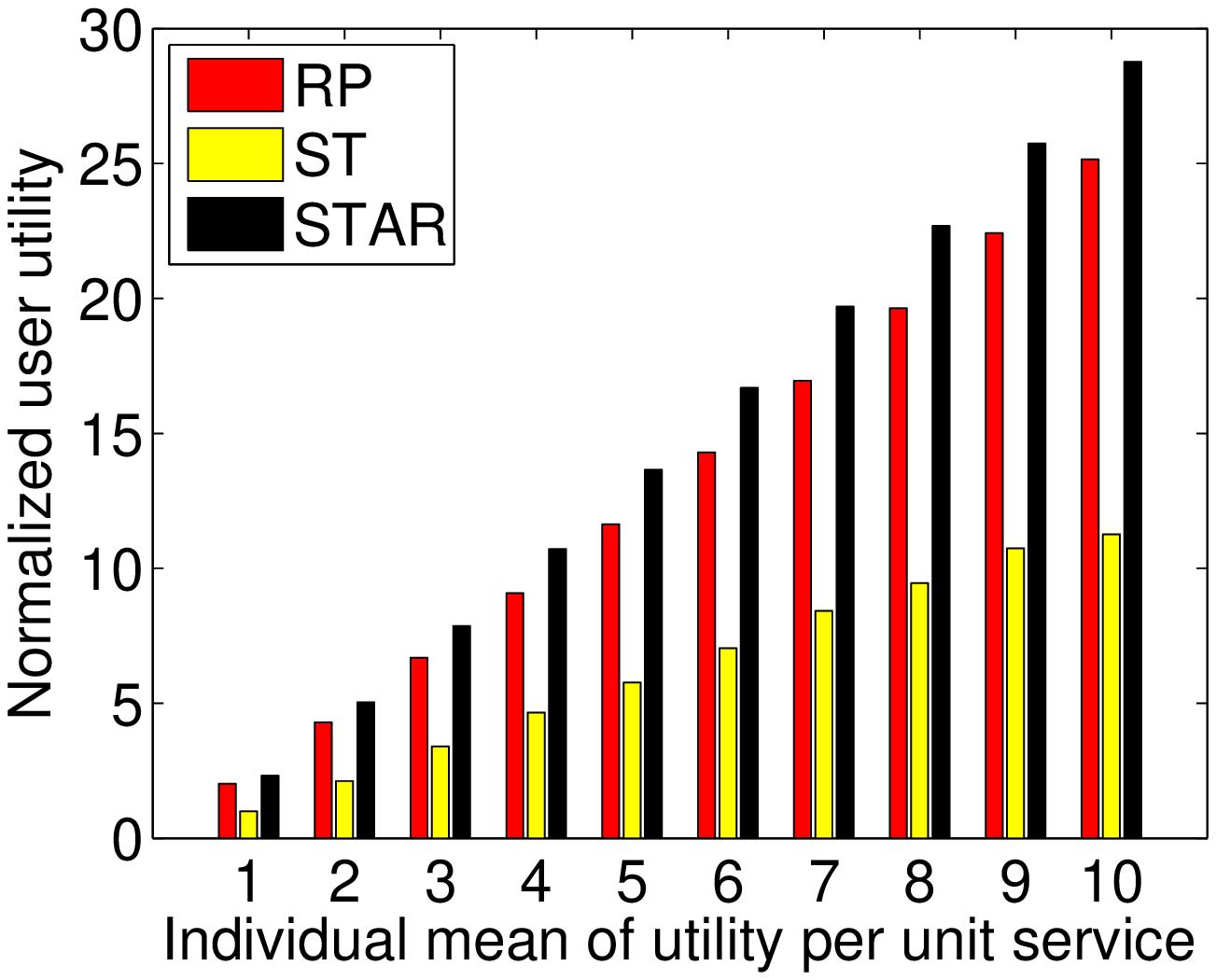}
\caption{Impact of $\mu_{U_i}$ on user service utility for the random setting.}
\label{fg:user_utility_Cw}
\end{minipage}
\end{figure*}

\subsection{Simulation Setup}

To illustrate the impact of different parameters of the mobile social network on the performance, we consider a random setting as follows. We simulate the social graph $G^S$ and the request
graph $G^R$ using the Erd\H{o}s-R\'{e}nyi (ER) graph model~\cite{Erdos60}, where a social edge and a request edge exist from one node to another with probability $P_S$ and $P_R$,
respectively. We assume that service is divisible. If a social edge exists, its social credit limit follows a normal distribution $N(\mu_S,\sigma^2_S)$, where $\mu_S$ and $\sigma^2_S$ denote
the mean and variance, respectively; if a physical edge exists, the amount of requested service and the utility per unit service follows a normal distribution $N(\mu_R,\sigma^2_R)$ and
$N(\mu_U,\sigma^2_U)$ respectively. We set default parameter values as: $N = 10$, $P_S = 0.2$, $P_R = 0.2$, $\mu_S = \mu_R = 5$, $\sigma^2_S = \sigma^2_R = 1$, $\mu_U = 10$, $\sigma^2_U=2$.


To evaluate the performance of the STAR mechanism in practice, we also consider a practical setting. Specifically, we generate the social graph according to the real dataset from
Brightkite~\cite{Brightkite}. Brightkite is a online social networking service based on mobile phones where users share their checking-in locations in an explicit social network. For this
dataset, we illustrate the social network structure of 20 users in Fig.~\ref{fg:real_social_graph}(a) and the users' degree of social edge in Fig.~\ref{fg:real_social_graph}(b). We also plot
the average number of social edge between a pair of users (in analogy to the probability of social edge in the ER model) versus the number of users in Fig.~\ref{fg:real_Ps_N}(a), and plot the average degree of social edge in Fig.~\ref{fg:real_Ps_N}(b). We simulate the request graph based on the context of spectrum crowdsensing discussed in Section~\ref{sc:model:example}. We
consider 5 licensed transmitters and $N$ users randomly located in a $1000m \times 1000m$ area. The licensed transmitters operate on 5 orthogonal channels, respectively. We assume that the
utility of a user's sensing service for a channel is equal to the inverse of its distance from the licensed transmitter that operates on that channel. Each user randomly selects one channel,
and requests sensing service for that channel from at most 3 users randomly selected from the other users who have better channel conditions than itself for that channel. We assume that the
sensing service is indivisible. The social credit limit and the service request amount are randomly drawn from $\{1,\cdots, N_S\}$ and $\{1,\cdots,N_R\}$, respectively. We set $N_S = N_R = 5$ as default values.


\subsection{Simulation Results}

\subsubsection{System Efficiency}

\begin{figure*}[t]
\begin{minipage}{.33\textwidth}
\vspace{-0.4cm}
\includegraphics[width=1\textwidth]{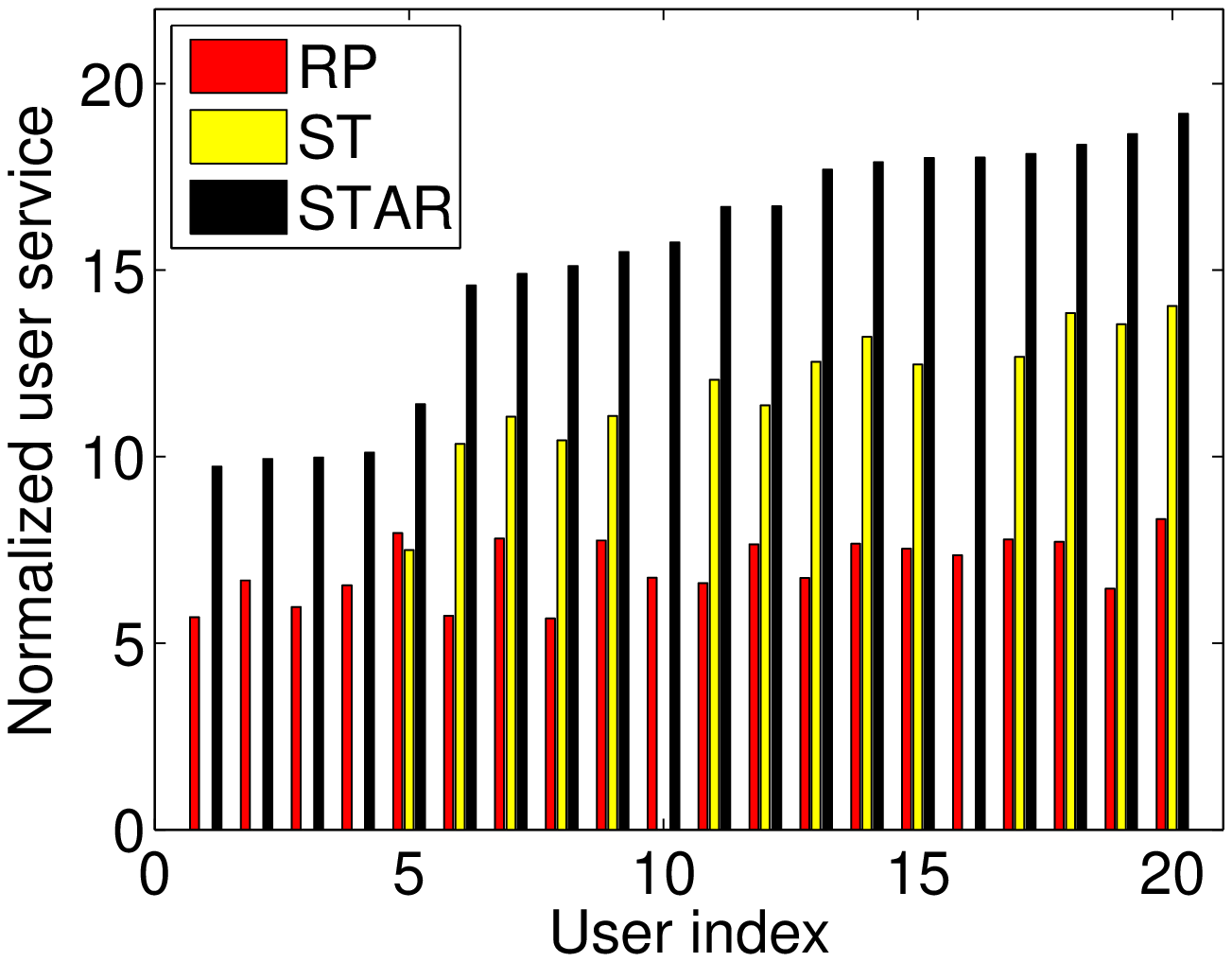}
\caption{Individual user service amount for the practical setting.}
\label{fg:real_user_service_Ps}
\end{minipage}
\begin{minipage}{.33\textwidth}
\includegraphics[width=1\textwidth]{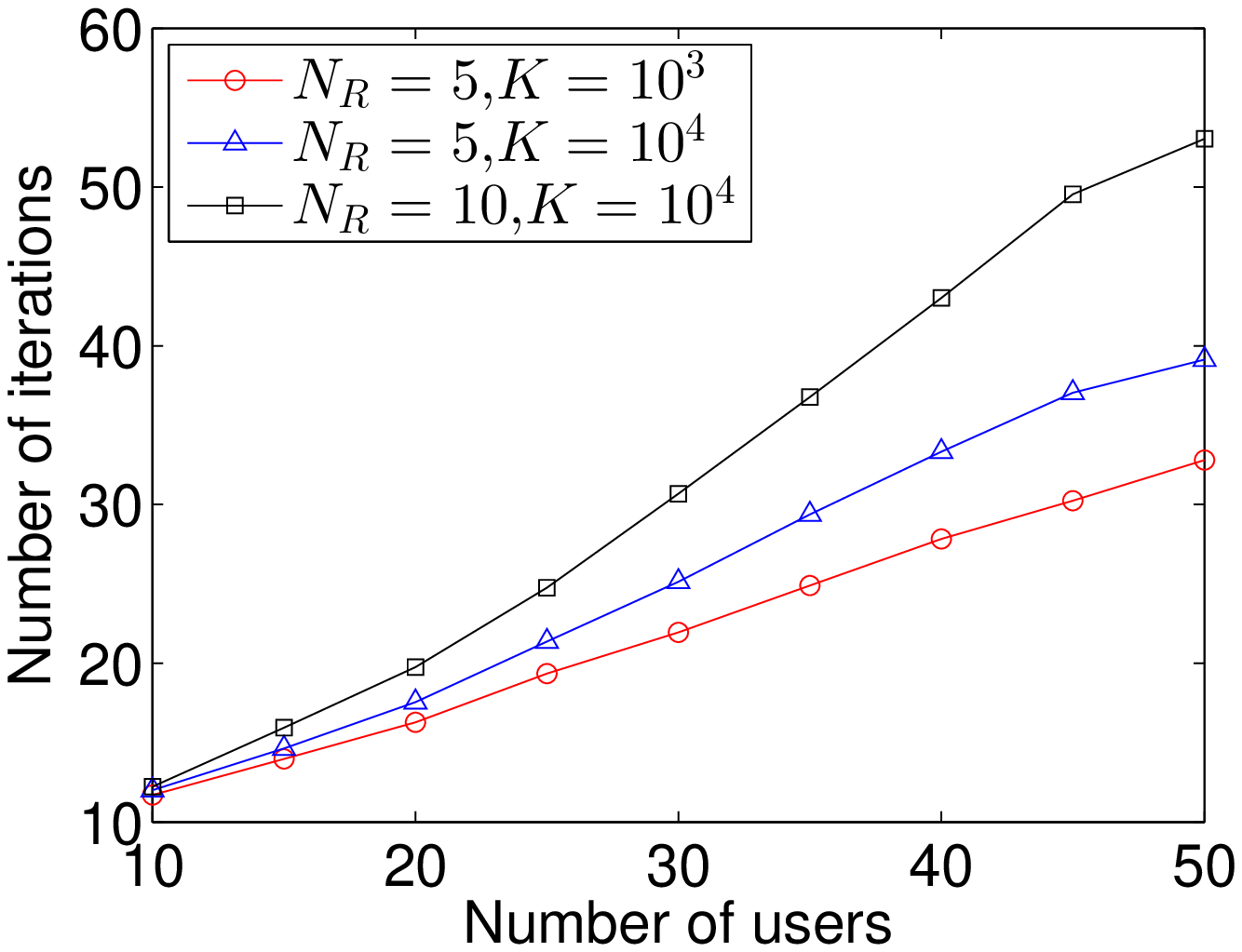}
\caption{Number of cycle-canceling iterations in Algorithm \ref{al:cycle_canceling} vs. number of users for the practical setting.}
\label{fg:real_iteration_N}
\end{minipage}
\begin{minipage}{.33\textwidth}
\vspace{-0.4cm}
\includegraphics[width=1\textwidth]{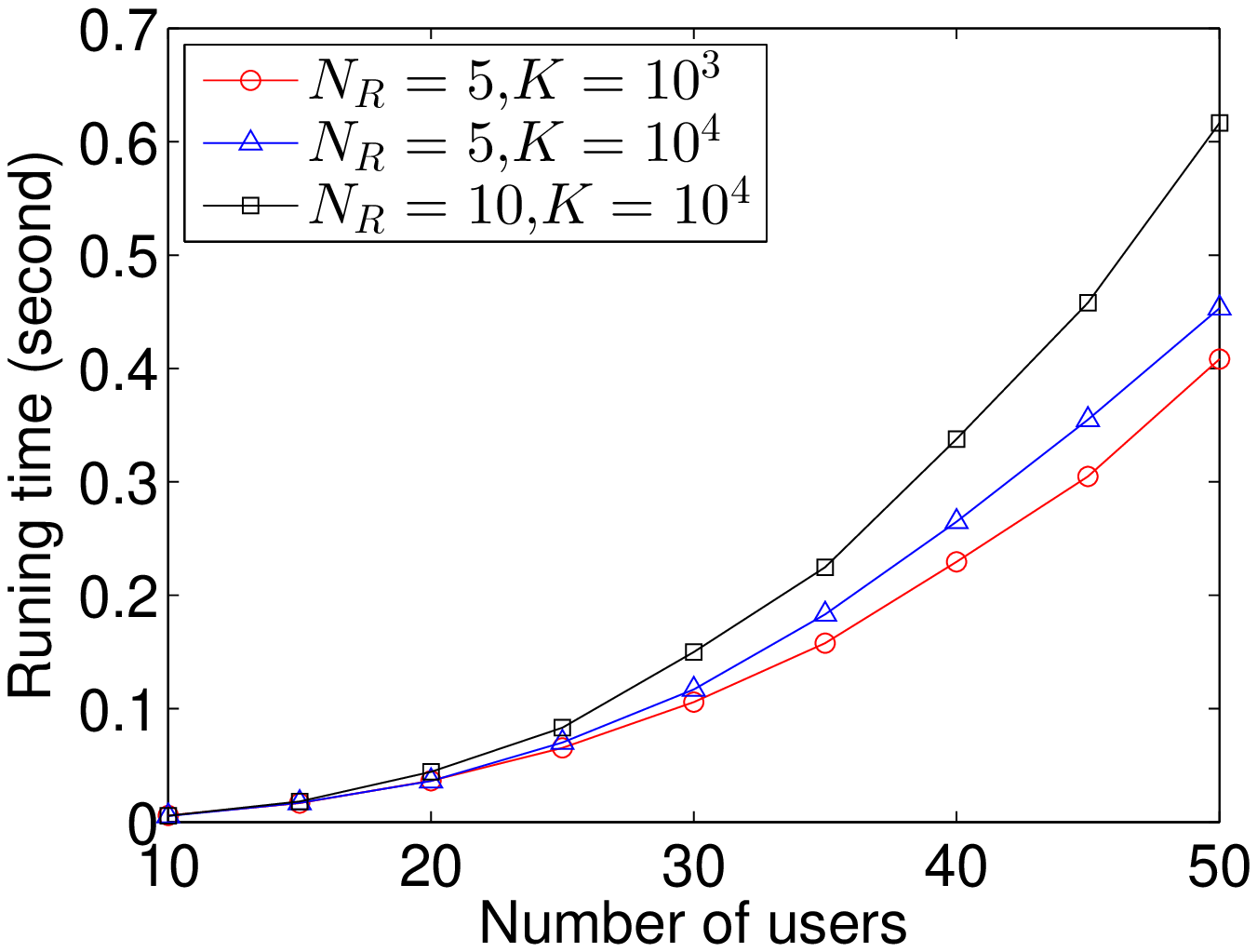}
\caption{Running time of Algorithm \ref{al:cycle_canceling} vs. number of users for the practical setting.}
\label{fg:real_time_N}
\end{minipage}
\end{figure*}

We compare the system performance of the STAR mechanism with the benchmark mechanisms RP and ST. We first evaluate the maximum total amount of service provided under different mechanisms. To
highlight the performance comparison, we normalize the results of STAR and ST with respect to RP. We illustrate the impact of $P_S$, $P_R$, $\mu_S$, $\mu_R$, and $N$ on the maximum total
amount of provided service in Figs. \ref{fg:total_service_Ps}-\ref{fg:total_service_N}, respectively. As expected, the performance of STAR always dominates that of RP and ST, which is due to
that STAR jointly exploits social trust and reciprocity. Figs. \ref{fg:total_service_Ps} and \ref{fg:total_service_Cs} show that STAR and ST perform better with respect to RP as $P_S$ or
$\mu_S$ increases. This is because that as social trust improves, more service can be provided using social trust under STAR and ST, while RP does not benefit from the improved social trust.
On the other hand, Figs. \ref{fg:total_service_Pd} and \ref{fg:total_service_Cd} show that STAR and ST perform worse with respect to RP as $P_R$ or $\mu_R$ increases. The reason is that as
users have more service requests among each other, a significant part of the increment in service request can be satisfied using reciprocity. We observe from Figs. \ref{fg:total_service_Ps}
and \ref{fg:total_service_Cs} that the performance gap between STAR and ST decreases as $P_S$ increases, while it remains almost the same as $\mu_S$ increases. This shows that the
connectivity of the social network has a greater impact on the performance of ST than the social trust levels. Due to this reason, Fig. \ref{fg:total_service_N} shows that the performance gap
between STAR and ST decreases as $N$ increases, since the connectivity of the social network  improves as the number of users increases. We also evaluate the maximum total utility of service
provided under different mechanisms. We illustrate the impact of $\mu_U$ on the maximum total utility of provided service in Fig. \ref{fg:total_utility_Cw}. We observe that the performance
gaps among RP, ST, and STAR remain almost the same as $\mu_U$ increases. This is as expected, since the utility per unit service acts as a ``scaling'' factor that has the same effect on the
performance of different mechanisms.

Next we evaluate the \emph{request completion ratio} under different mechanisms, which is defined as the ratio of the amount of provided service to the amount of requested service. We
illustrate the impact of $P_R$, $\mu_R$, and $N$ on the request completion ratio in Figs. \ref{fg:total_ratio_Pd}-\ref{fg:total_ratio_N}, respectively. We observe from Figs.
\ref{fg:total_ratio_Pd} and \ref{fg:total_ratio_Cd} that for all mechanisms, the total amount of provided service increases faster than that of requested service as $P_R$ increases, while it
increases slower as $\mu_R$ increases. This shows that a large diversity of users' service requests is beneficial for system efficiency. Due to this reason, as illustrated in Fig.
\ref{fg:total_ratio_N}, the request completion ratio improves for all mechanisms as the number of users increases.

For the practical setting, Figs. \ref{fg:real_service_N}-\ref{fg:real_ratio_N} illustrate the total service amount, total service utility, and request completion ratio when the total utility
of provided service is maximized, respectively, as $N$ increases. We can see that STAR always significantly outperforms RP and ST, with a performance gain ranging from 14\% to 82\%,
especially when the number of users is small.

\subsubsection{Individual Performance}

We evaluate individual users' performance under different mechanisms when the system efficiency is maximized. To demonstrate the impact of a particular parameter, we vary that parameter for
different users, while keeping other parameters the same for all users. We also normalize the results to highlight the performance comparison. In Figs.
\ref{fg:user_service_Ps}-\ref{fg:user_service_Cd}, we illustrate the amount of received service of each user (i.e., the amount of satisfied service requests of each user) for a system of 10
users when the total amount of provided service is maximized, where users are different only in a user's probability of having social edge from another user $P_{S_i}$, probability of having
service request from another user $P_{R_i}$, mean of social credit limit from another user $\mu_{S_i}$, and mean of service request amount from another user $\mu_{R_i}$, respectively. In
Figs. \ref{fg:user_service_Cw}-\ref{fg:user_utility_Cw}, we illustrate each user's received service amount and received service utility when the total utility of provided service is
maximized, where users are different only in a user's utility per unit service $\mu_{U_i}$. We observe that each user always performs better under STAR than under RP and ST. This shows that
STAR can improve each individual user's performance while the system objective is to maximize system efficiency. We also observe that an individual user performs better than other users if it
has a larger parameter value than others. This shows that STAR can achieve \emph{service differentiation}, which is a desirable property for fairness: if a user has more social trust or
service requests from others than other users have, then that user can also receive more service than others.

Fig.~\ref{fg:real_user_service_Ps} illustrates each individual user's received service amount for the 20 users in the real dataset \cite{Brightkite} with the social network structure as given
in Fig.~\ref{fg:real_social_graph}. We observe from Fig.~\ref{fg:real_social_graph}(b) that the degree of social edge can be very different for different users in real social networks.
Accordingly, Fig.~\ref{fg:real_user_service_Ps} shows that users with higher degrees (with larger user indices) receive more service than those with lower degrees.



%

\subsubsection{Computational Complexity}

We evaluate the computational complexity of using Algorithm \ref{al:cycle_canceling} to find the maximum total utility of provided service under the STAR mechanism for the practical setting.
We convert the service utility values into integers by multiplying them by a large integer $K$ and rounding them down to the respective nearest integers. We run simulations on a Windows 7
desktop with 3.1GHz CPU and 8GB memory. We illustrate the number of cycle-canceling iterations (i.e., the iterations of the \textbf{while} loop) and the running time of executing Algorithm
\ref{al:cycle_canceling} as $N$ increases for different values of $N_R$ and $K$ in Fig. \ref{fg:real_iteration_N} and Fig.\ref{fg:real_time_N}, respectively. We assume that $N_S=N_R$. We
observe that the number of iterations increases almost linearly in the number of users while the running time is increasing quadratically. This shows that Algorithm \ref{al:cycle_canceling}
is scalable for large systems in practice. As expected, we also observe that the computational complexity is higher when $N_R$ or $K$ is larger.

\section{Related Work}\label{sc:related}

There have been numerous studies on incentive design for stimulating user cooperation in networks. Existing literature on this subject can be broadly classified into three categories. One category of work makes use of reciprocity (also known as \emph{barter})~\cite{Yuen03,Ganesan05,Shevade08}. Although a reciprocity-based approach is simple to implement, it is inefficient in general since synchronously matched requests are unusual. Another category is based on (virtual) currency~\cite{Wang06,Zhong07,Zhang11,Yang12,Koutsopoulos13}, in which a user earns currency by providing service to others and spends currency to receive service from others. The use of currency as a medium of exchange overcomes the shortcoming of reciprocity-based approaches by enabling users to ``asynchronously trade'' service. However, a major drawback of using currency is that it incurs a high implementation overhead, mainly due to the need to inhibit malicious manipulation among users without mutual trust. Consider, for example, \emph{Bitcoin}~\cite{Bitcoin} which has recently drawn widespread attention as a promising digital currency. The creation and transfer of bitcoins need to consume considerable computing resources so that they can be secured against potential cheating using cryptographic tools. Reputation-based approaches~\cite{Marti00,Buchegger02,Jaramillo07} constitute the third category. Since reputation score can be viewed as a form of currency, these approaches share the same
advantages and disadvantages as the currency-based ones.

The social credit model used in this paper falls into the class of \emph{credit networks}~\cite{DeFigueiredo05,Dandekar10,Dandekar12,Liu10}. The credit is similar to a currency in
that there is a need to keep track of the credit information between each pair of neighbor users in the credit network. However, since the credit is ensured by existing trust among users, it
obviates the need to secure the credit against cheating, and therefore can reduce implementation overhead significantly.


Compared to the studies mentioned above, the STAR mechanism overcomes the inefficiency of only using reciprocity by using social credit as a ``local'' currency, while it also circumvents the
high implementation overhead incurred by a currency-based approach since social credit is ``secured'' by existing social trust. Therefore, STAR can efficiently stimulate users to provide
service in a cost-effective way.

Exploiting social aspect for mobile networking is an emerging paradigm for network design and optimization \cite{Chen13,Chen14}. Very few work have exploited both social trust
and reciprocity for stimulating cooperation in networks. \cite{Chen13} has recently studied using social trust and reciprocity to stimulate cooperative communication based on a coalitional
game. Our work is different from~\cite{Chen13} in several ways including that each user in the latter can participate in \emph{at most one} reciprocity cycle and social trust levels are \emph{unlimited} therein.


\section{Conclusion}\label{sc:conclusion}

In this paper, we have proposed a socially-aware crowdsensing system that exploits social trust to stimulate users' participation. The incurred implementation overhead is low since it
obviates the need of a global currency. For this system, we have designed STAR, an incentive mechanism using a synergistic marriage of social trust and reciprocity. Based on the STAR
mechanism, we have shown that all sensing requests can be satisfied if and only if users who request more sensing service than they can provide can transfer sufficient social credit to users
who can provide more than they request. We have also developed an efficient algorithm to maximize the utility of sensing service provided under STAR, for both cases of divisible and
indivisible service. Extensive simulation results have confirmed that STAR can achieve significantly better efficacy than using social trust only or reciprocity only.


\bibliographystyle{IEEEtran}
\bibliography{IEEEabrv,social_reciprocity_cooperation_Bib}

\begin{thebibliography}{10}
\providecommand{\url}[1]{#1}
\csname url@samestyle\endcsname
\providecommand{\newblock}{\relax}
\providecommand{\bibinfo}[2]{#2}
\providecommand{\BIBentrySTDinterwordspacing}{\spaceskip=0pt\relax}
\providecommand{\BIBentryALTinterwordstretchfactor}{4}
\providecommand{\BIBentryALTinterwordspacing}{\spaceskip=\fontdimen2\font plus
\BIBentryALTinterwordstretchfactor\fontdimen3\font minus
  \fontdimen4\font\relax}
\providecommand{\BIBforeignlanguage}[2]{{%
\expandafter\ifx\csname l@#1\endcsname\relax
\typeout{** WARNING: IEEEtran.bst: No hyphenation pattern has been}%
\typeout{** loaded for the language `#1'. Using the pattern for}%
\typeout{** the default language instead.}%
\else
\language=\csname l@#1\endcsname
\fi
#2}}
\providecommand{\BIBdecl}{\relax}
\BIBdecl

\bibitem{Gong14GlobalSIP}
X.~Gong, X.~Chen, J.~Zhang, and H.~V. Poor, ``From social trust assisted
  reciprocity {(STAR)} to utility-optimal crowdsensing in mobile networks,'' in
  \emph{Proc. IEEE GlobalSIP 2014}.

\bibitem{gartner}
\BIBentryALTinterwordspacing
``Gartner: Worldwide {PC}, tablet and mobile phone shipments to grow 4.5
  percent in 2013.'' [Online]. Available:
  \url{http://www.gartner.com/newsroom/id/2610015}
\BIBentrySTDinterwordspacing

\bibitem{Yang12}
D.~Yang, G.~Xue, X.~Fang, and J.~Tang, ``Crowdsourcing to smartphones:
  incentive mechanism design for mobile phone sensing,'' in \emph{Proc. ACM
  MOBICOM 2012}.

\bibitem{Luo12}
T.~Luo and C.-K. Tham, ``Fairness and social welfare in incentivizing
  participatory sensing,'' in \emph{Proc. IEEE SECON 2012}.

\bibitem{Koutsopoulos13}
I.~Koutsopoulos, ``Optimal incentive-driven design of participatory sensing
  systems,'' in \emph{Proc. IEEE INFOCOM 2013}.

\bibitem{emarketer}
\BIBentryALTinterwordspacing
``e{M}arketer: Social networking reaches nearly one in four around the world.''
  [Online]. Available:
  \url{http://www.emarketer.com/Article/Social-Networking-Reaches-Nearly-One-Four-Around-World/1009976}
\BIBentrySTDinterwordspacing

\bibitem{waze}
\BIBentryALTinterwordspacing
``Waze: outsmarting traffic, together.'' [Online]. Available:
  \url{https://www.waze.com/}
\BIBentrySTDinterwordspacing

\bibitem{Sahai09}
e.~a. Sahai, A., ``Cognitive radios for spectrum sharing,'' \emph{IEEE Signal
  Processing Magazine}, vol.~26, pp. 140--145, Jan. 2009.

\bibitem{Liu10}
Z.~Liu, H.~Hu, Y.~Liu, K.~W. Ross, Y.~Wang, and M.~Mobius, ``{P2P} trading in
  social networks: The value of staying connected,'' in \emph{Proc. IEEE
  INFOCOM 2010}.

\bibitem{Ahuja93}
R.~Ahuja, T.~Magnanti, and J.~Orlin, \emph{Network flows: Theory, algorithms,
  and applications}.\hskip 1em plus 0.5em minus 0.4em\relax Prentice Hall,
  1993.

\bibitem{Huang06}
X.~Huang, ``Negative-weight cycle algorithms,'' in \emph{Proc. 2006
  International Conference on Foundations of Computer Science}.

\bibitem{Brightkite}
\BIBentryALTinterwordspacing
``{SNAP: Network datasets: Brightkite}.'' [Online]. Available:
  \url{http://snap.stanford.edu/data/loc-brightkite.html}
\BIBentrySTDinterwordspacing

\bibitem{Erdos60}
P.~Erdos and A.~Renyi, ``On the evolution of random graphs,''
  \emph{Publications of the Mathematical Institute of the Hungarian Academy of
  Sciences}, pp. 17--61, 1960.

\bibitem{Yuen03}
W.~H. Yuen, R.~D. Yates, and S.-C. Mau, ``Exploiting data diversity and
  multiuser diversity in non-cooperative mobile infostation networks,'' in
  \emph{Proc. IEEE INFOCOM 2003}.

\bibitem{Ganesan05}
P.~Ganesan and M.~Seshadri, ``On cooperative content distribution and the price
  of barter,'' in \emph{Proc. IEEE ICDCS 2005}.

\bibitem{Shevade08}
U.~Shevade, H.~H. Song, L.~Qiu, and Y.~Zhang, ``Incentive-aware routing in
  {DTNs},'' in \emph{Proc. IEEE ICNP 2008}.

\bibitem{Wang06}
W.~Wang, S.~Eidenbenz, Y.~Wang, and X.-Y. Li, ``{OURS}: optimal unicast routing
  systems in non-cooperative wireless networks,'' in \emph{Proc. ACM MOBICOM
  2006}.

\bibitem{Zhong07}
S.~Zhong and F.~Wu, ``On designing collusion-resistant routing schemes for
  non-cooperative wireless ad hoc networks,'' in \emph{Proc. ACM MOBICOM 2007}.

\bibitem{Zhang11}
C.~Zhang, X.~Zhu, Y.~Song, and Y.~Fang, ``C4: A new paradigm for providing
  incentives in multi-hop wireless networks,'' in \emph{Proc. IEEE INFOCOM
  2011}.

\bibitem{Bitcoin}
\BIBentryALTinterwordspacing
``Bitcoin: An open source {P2P} digital currency.'' [Online]. Available:
  \url{http://bitcoin.org/en/}
\BIBentrySTDinterwordspacing

\bibitem{Marti00}
S.~Marti, T.~J. Giuli, K.~Lai, and M.~Baker, ``Mitigating routing misbehavior
  in mobile ad hoc networks,'' in \emph{Proc. ACM MOBICOM 2000}.

\bibitem{Buchegger02}
S.~Buchegger and J.-Y.~L. Boudec, ``Performance analysis of the {CONFIDANT}
  protocol,'' in \emph{Proc. ACM MOBIHOC 2002}.

\bibitem{Jaramillo07}
J.~Jaramillo and R.~Srikant, ``{DARWIN}: Distributed and adaptive reputation
  mechanism for wireless networks,'' in \emph{Proc. ACM MOBICOM 2007}.

\bibitem{DeFigueiredo05}
D.~B. DeFigueiredo and E.~T. Barr, ``Trustdavis: A non-exploitable online
  reputation system,'' in \emph{Proc. IEEE E-Commerce Technology 2005}.

\bibitem{Dandekar10}
P.~Dandekar, A.~Goel, R.~Govindan, and I.~Post, ``Liquidity in credit networks:
  A little trust goes a long way,'' in \emph{Proc. ACM EC 2011}.

\bibitem{Dandekar12}
P.~Dandekar, A.~Goel, M.~P. Wellman, and B.~Wiedenbeck, ``Strategic formation
  of credit networks,'' in \emph{Proc. ACM WWW 2012}.

\bibitem{Chen13}
X.~Chen, B.~Proulx, X.~Gong, and J.~Zhang, ``Social trust and social
  reciprocity based cooperative {D2D} communications,'' in \emph{Proc. ACM
  MOBIHOC 2013}.

\bibitem{Chen14}
X.~Chen, X.~Gong, L.~Yang, and J.~Zhang, ``A social group utility maximization
  framework with applications in database assisted spectrum access,'' in
  \emph{Proc. IEEE INFOCOM 2014}.

\end{thebibliography}

\end{document}